\documentclass{article}
\usepackage{graphicx,times}
\begin {document}
\pagenumbering{arabic}
\pagestyle {plain}
\frenchspacing
\parindent 1.0 cm
\parskip 0.6cm
\vspace*{1.0 cm}
\begin{flushleft}
{\bf\large
Energy levels, radiative rates,  and lifetimes  for transitions in W XL} \vspace{0.5 cm}
\\{\sf Kanti M. Aggarwal} and {\sf Francis P. Keenan}\\ \vspace*{0.3 cm} 
Astrophysics Research Centre, School of Mathematics and Physics, Queen's University Belfast,\\Belfast BT7 1NN,
Northern Ireland, UK.\\  \vspace*{0.1 cm}

\vspace*{0.2 cm}  
{\bf ABSTRACT} \\ 

Energy levels and radiative rates are reported for transitions in Br-like W XL,  calculated with the general-purpose relativistic atomic structure package ({\sc grasp}). Configuration interaction (CI) has been included among 46 configurations (generating 4215 levels) over a wide energy range up to 213 Ryd. However, for conciseness results are only listed for the lowest 360 levels (with energies up to $\sim$ 43 Ryd), which mainly belong to the  4s$^2$4p$^5$, 4s$^2$4p$^4$4d, 4s$^2$4p$^4$4f,  4s4p$^6$,  4p$^6$4d, 4s4p$^5$4d, 4s$^2$4p$^3$4d$^2$, and 4s$^2$4p$^3$4d4f   configurations, and provided for four types of transitions, i.e. E1, E2, M1, and M2. Comparisons are made with existing (but limited) results. However, to fully assess the accuracy of our data, analogous calculations have been performed with the  flexible atomic code ({\sc fac}), including even a larger CI than in {\sc grasp}.  Our energy levels are estimated to be accurate to better than  0.02 Ryd, whereas results for radiative rates (and lifetimes) should be  accurate to  better than 20\% for a majority of the strong transitions. 

\vspace*{0.1 cm}  
Received December 13, 2013; accepted February 5, 2014\\ 
------------------------------------------------------------------------------------------------------------------------------------------------ \\
\vspace*{0.2 cm}
{\bf Running Title}: {\em K. M. Aggarwal and F. P. Keenan / Atomic Data and Nuclear Data Tables xxx (2014) xxx-xxx}
\end{flushleft}
\newpage

\begin{flushleft}
{\bf Contents}

\begin{tabular}{llr}
1.   &  Introduction  .........................................                &   00 \\
2.   &  Energy levels  ........................................            &   00 \\
3.   &  Radiative rates  ....................................           &   00 \\
4.   &  Lifetimes  .........................................                      &   00 \\
5.   &  Conclusions  ....................................		        &   00 \\
     &  Acknowledgments  .............................                   &   00 \\
     &  References  ........................................		         &   00 \\

\end{tabular}
\end{flushleft}
\begin{flushleft}
   Explanation of Tables \\ \vspace*{0.2 cm}   
   Tables\\ \vspace*{0.2 cm} 

\begin{tabular}{rlr}
                                                                                                                     
 1. & Configurations and  levels of W XL.  .................................................         &    00 \\
 2. &  Energies (Ryd) for the lowest 360 levels of W XL and their  lifetimes (in s). .................................................        &    00 \\
 3. & Transition wavelengths ($\lambda_{ij}$ in ${\rm \AA}$), radiative rates ($A_{ji}$ in s$^{-1}$), oscillator strengths  .................................... &    00 \\
     & (f$_{ij}$, dimensionless), line strengths S (in atomic unit) for electric dipole (E1), and ............................................. &    00 \\
     & A$_{ji}$ for electric quadrupole (E2), magnetic dipole (M1), and magnetic quadrupole (M2) transitions in W XL. .....                   &    00 \\ 
\end{tabular}	
\end{flushleft}															      

\newpage
\begin{flushleft}
{\bf 1. Introduction}
\end{flushleft}

\parindent = 1 cm 

Tungsten (W) is an element of active research for studies of fusion plasmas. Since it is an important constituent of tokamak reactor walls, its importance has further increased with  the developing  ITER project. Due to the high temperatures of fusion plasmas, atomic data  (including energy levels and  radiative decay rates) are required for many of W ions, mainly to assess  and control the radiation loss. Hence there have been several theoretical studies of W ions -- see for example, Fournier \cite{kbf} and Jonauskas et al. \cite{jgk}. Similarly, there have been laboratory measurements of W emission lines for many of its ions   -- see for example, Utter et al. \cite{sbu} and Clementson et al. \cite{ll1}.  These data, including for W XL,  have been  compiled by NIST (National Institute of Standards and Technology), and are  available at their  website {\tt http://physics.nist.gov/PhysRefData/ASD/levels\_form.html}.

Compared to other W ions there has been less focus on Br-like W XL. In fact, the only calculation available is by Fournier \cite{kbf}, who reported energies for 29 levels of the 4s$^2$4p$^5$, 4s$^2$4p$^4$4d, and 4s4p$^6$ configurations, and a selected few of 4s$^2$4p$^4$4f, 4s$^2$4p$^4$5d, and 3d$^9$4s$^2$4p$^6$. Similarly, his data for radiative rates (A- values) are limited to a few transitions, while for plasma modelling one normally requires a complete set of A- values for {\em all} transitions. 

W XL is a heavy ion (Z = 74) and hence the inclusion of  relativistic effects is very important to accurately determine energy levels and subsequently other  parameters. However, for this ion configuration interaction (CI) is equally important. For this reason, Fournier \cite{kbf} included CI among 10 even and 10 odd configurations  --  see his Table 1.  However, he omitted some important configurations, such as 4s$^2$4p$^3$4d$^2$, which  generate 141 odd parity levels in the 22--42 Ryd energy range  -- see Table 1 of Aggarwal and Keenan \cite{cjp}. Our {\em aim} is to improve upon these earlier calculations and provide  a complete set of atomic data among a larger number of levels. 

Before we discuss our work, we note that recently S. Aggarwal et al.  \cite{ajm} calculated  energy levels, oscillator strengths, radiative rates,  and lifetimes for 33 levels of W XL,  belonging to the 4s$^2$4p$^5$, 4s$^2$4p$^4$4d, 4s4p$^6$, and 3d$^9$4s$^2$4p$^6$ configurations. However, their results are unreliable, with  discrepancies in A- values and lifetimes with the present calculations of  up to two and four orders of magnitude, respectively, as shown  by Aggarwal and Keenan \cite{cjp}.  Therefore, we do not discuss their results further.

 For the calculations, we have adopted the {\sc grasp} (general-purpose relativistic atomic structure package) code,  originally developed by Grant  et al.  \cite{grasp0}, and significantly revised by one of its authors (Dr. P. H. Norrington). This code is referred to as  GRASP0 and is available at  {\tt http://web.am.qub.ac.uk/DARC/}. It is fully relativistic,  based on the $jj$ coupling scheme, and includes higher-order relativistic corrections arising from the Breit interaction and QED (quantum electrodynamics) effects. Results obtained from this code are comparable with other subsequent  versions, such as GRASP2K \cite{grasp2k}--\cite{grasp2kk}.

\begin{flushleft}
{\bf 2. Energy levels}
\end{flushleft}
\parindent = 1 cm 

For our calculations we have used the option of {\em extended average level} (EAL),  in which a weighted (proportional to 2$j$+1) trace of the Hamiltonian matrix is minimised. This produces a compromise set of orbitals describing closely-lying states with moderate accuracy, and yields results comparable to other options, such as {\em average level} (AL) -- see for example,  Aggarwal  et al.  for several ions of Kr \cite{kr} and Xe \cite{xe}.  Since CI is very important for W XL as already noted, we have performed a series of calculations with increasing amount of CI with up to  42 configurations, namely 4s$^2$4p$^5$, 4s$^2$4p$^4$4d, 4s$^2$4p$^4$4f,  4s4p$^6$,  3d$^9$4s$^2$4p$^6$,   4p$^6$4d/4f, 4s4p$^5$4d/4f, 4p$^3$4d$^2$/4f$^2$/4d4f, 4s$^2$4p$^2$4d$^3$, 4s$^2$4p4d$^4$, 4s$^2$4p$^2$4d$^2$4f, 4s4p$^3$4d$^3$, 4p$^5$4d$^2$, 3d$^9$4s$^2$4p$^5$4d/4f, 3p$^5$3d$^{10}$4s$^2$4p$^6$,  3p$^5$3d$^{10}$4s$^2$4p$^5$4d/4f, 4s4p$^5$5$\ell$, 4p$^6$5$\ell$, 4s$^2$4p$^4$5$\ell$, and 3d$^9$4s$^2$4p$^5$5$\ell$. These generate a total of 4128 levels  and are spread over a wide energy range up to 213 Ryd. Many of these configurations have overlapping energy levels, and hence strongly interact and intermix. The levels each configuration generates and their energy ranges are listed in Table 1. Also listed in the table are the configurations included by Fournier \cite{kbf}.

The importance of CI on energy levels and A- values has been clearly demonstrated in our earlier work \cite{cjp}. Differences of up to 0.4 Ryd with the energy levels of Fournier \cite{kbf} were noted  for several levels -- see Table 2 of \cite{cjp}, mainly due to the larger CI included in our calculations. Discrepancies for A- values were larger, up to a factor of three, for several transitions -- see Table 3 of \cite{cjp}. However, in  \cite{cjp}  results were only provided for the lowest 31 levels of the 4s$^2$4p$^5$, 4s$^2$4p$^4$4d, and 4s4p$^6$ configurations, and two of 3d$^9$4s$^2$4p$^6$, which are common with those of \cite{kbf}. Similarly, A- values were only reported from the lowest two levels of the  4s$^2$4p$^5$  ground configuration. Here we list  data for a larger number of levels, up to 360, and in addition four more configurations have been included, apart from the earlier 42 listed above. These are: (4s$^2$4p$^4$) 6s, 6p, 6d and 6f, which generate 87 levels in the 81-106 Ryd energy range and overlap with  many of the other (1--42) configurations listed in Table 1. However, we note that the effect of including these additional four configurations is {\em insignificant} on the energy levels and A- values reported in our earlier work \cite{cjp}. 

The 46 configurations of W XL listed in Table 1 generate 4215 levels in total, but for conciseness energies are listed in Table 2 for only the lowest 360 levels (with E up to $\sim$ 43 Ryd), which mostly belong to the  4s$^2$4p$^5$, 4s$^2$4p$^4$4d, 4s$^2$4p$^4$4f,  4s4p$^6$,  4p$^6$4d, 4s4p$^5$4d, 4s$^2$4p$^3$4d$^2$, and 4s$^2$4p$^3$4d4f configurations. Energies for all levels may be obtained electronically on request from one of the authors (KMA: K.Aggarwal@qub.ac.uk). 

For ions where CI is very important, it is very difficult to assign a unique label for each level, because many of these mix strongly -- see for example Ti VI \cite{tivi} and Ti VII \cite{tivii}. In Table A we list the mixing coefficients for 31 levels of the 4s$^2$4p$^5$, 4s$^2$4$^4$4d, and 4s4p$^6$ configurations, and except for the lowest two,  almost all others are well mixed. Consequently,  some of the levels can be easily interchanged, such as 7/60 (4s$^2$4p$^4$($^1$S)4d   $^2$D$   _{3/2 }$ and 4s$^2$4p$^4$($^3$P)4d  $^4$F$   _{3/2 }$), 13/26 (4s$^2$4p$^4$($^1$D)4d  $^2$F$   _{5/2 }$ and 4s$^2$4p$^4$($^1$D)4d  $^2$D$   _{5/2 }$), and 16/22 (4s$^2$4p$^4$($^1$D)4d  $^2$D$   _{3/2 }$ and 4s$^2$4p$^4$($^3$P)4d $^2$P$   _{3/2 }$). Therefore, the $LSJ$ designations provided in Table 2 are mainly for guidance and should not be taken as definitive. In Table B we provide similar mixing coefficients for the remaining levels, but only up to 65.  These  are more strongly mixed than those in Table A -- see for  example 31 and 48, i.e. 4s$^2$4p$^3$($^2$P)4d$^2$($^3$F) $^4$D$^o _{1/2 }$ and 4s$^2$4p$^3$($^4$S)4d$^2$($^3$F)  $^6$F$^o _{1/2 }$.

Experimental energies for W XL (see the NIST website) are limited to only a few levels for which detailed comparisons have already been made in our earlier work \cite{cjp}. In the absence of other theoretical results (except  of \cite{kbf}), it is hence difficult to fully assess the accuracy of our energy levels. To make such an  assessment  we have therefore employed the  {\em Flexible Atomic Code} ({\sc fac}) of Gu \cite{fac},  available from the website {\tt http://sprg.ssl.berkeley.edu/$\sim$mfgu/fac/}. This is also a fully relativistic code which provides  results for energy levels and A- values comparable to {\sc grasp}, as already shown for several other ions, see for example:  Aggarwal  et al. for Kr \cite{kr} and Xe \cite{xe} ions, and the more recent work on Ti ions (\cite{tivi}--\cite{tivii} and \cite{tix}). A further  advantage of this code is its high efficiency which means that significantly large calculations can be performed within a reasonably short time. However, because of the strong mixing for several levels of W XL the identification of a level is comparatively more difficult in the FAC calculations than for GRASP.

As with {\sc grasp}, we have  performed a series of calculations  with the {\sc fac} code with increasing amount of CI,  but focus only on two. These are: (i) FAC1, which includes the same 4215 levels as in GRASP and (ii) FAC2, which also includes all possible combinations of the 4s$^2$4p$^3$5$\ell{\ell'}$ and 4s4p$^4$5$\ell{\ell'}$  configurations, i.e. 11,849 levels in total. Detailed comparisons for 33 levels of the 4s$^2$4p$^5$, 4s$^2$4p$^4$4d,  4s4p$^6$, and 3d$^9$4s$^2$4p$^6$ configurations have already been undertaken in our earlier paper \cite{cjp}, and  in Table C we provide a similar comparison for the remaining levels, but only up to 65. No measurements or other theoretical results are available for these levels.

For the levels listed in Table C, the energies obtained in our GRASP and FAC1 calculations are comparable (within 0.01 Ryd) and the orderings are also the same. Similarly, the energies in the FAC2 calculations closely agree with those of FAC1, but are higher by $\sim$ 0.01 Ryd for almost all levels. Therefore, the differences between the GRASP and FAC2 energies are less than 0.02 Ryd, and overall there is not much appreciable effect arising from  the extra CI included in the FAC2 calculations. Based on this comparison as well as the one shown in Table 2 of  \cite{cjp}, we can confidently state that our energy levels listed in Table 2 are accurate to $\sim$ 0.02 Ryd.

\begin{flushleft}
{\bf 3. Radiative rates}
\end{flushleft}

The absorption oscillator strength ($f_{ij}$), a dimensionless quantity,  and radiative rate A$_{ji}$ (in s$^{-1}$) for a transition $i \to j$ are related by the following expression:

\begin{equation}
f_{ij} = \frac{mc}{8{\pi}^2{e^2}}{\lambda^2_{ji}} \frac{{\omega}_j}{{\omega}_i}A_{ji}
 = 1.49 \times 10^{-16} \lambda^2_{ji} \frac{{\omega}_j}{{\omega}_i} A_{ji}
\end{equation}
where $m$ and $e$ are the electron mass and charge, respectively, $c$  the velocity of light,  $\lambda_{ji}$  the transition energy/wavelength in $\rm \AA$, and $\omega_i$ and $\omega_j$  the statistical weights of the lower $i$ and upper $j$ levels, respectively.
Similarly, the oscillator strength $f_{ij}$ (dimensionless) and the line strength $S$ (in atomic unit, 1 a.u. = 6.460$\times$10$^{-36}$ cm$^2$ esu$^2$) are related by the 
following standard equations:

\begin{flushleft}
for the electric dipole (E1) transitions: 
\end{flushleft} 
\begin{equation}
A_{ji} = \frac{2.0261\times{10^{18}}}{{{\omega}_j}\lambda^3_{ji}} S \hspace*{1.0 cm} {\rm and} \hspace*{1.0 cm} 
f_{ij} = \frac{303.75}{\lambda_{ji}\omega_i} S, \\
\end{equation}
\begin{flushleft}
for the magnetic dipole (M1) transitions:  
\end{flushleft}
\begin{equation}
A_{ji} = \frac{2.6974\times{10^{13}}}{{{\omega}_j}\lambda^3_{ji}} S \hspace*{1.0 cm} {\rm and} \hspace*{1.0 cm}
f_{ij} = \frac{4.044\times{10^{-3}}}{\lambda_{ji}\omega_i} S, \\
\end{equation}
\begin{flushleft}
for the electric quadrupole (E2) transitions: 
\end{flushleft}
\begin{equation}
A_{ji} = \frac{1.1199\times{10^{18}}}{{{\omega}_j}\lambda^5_{ji}} S \hspace*{1.0 cm} {\rm and} \hspace*{1.0 cm}
f_{ij} = \frac{167.89}{\lambda^3_{ji}\omega_i} S, 
\end{equation}

\begin{flushleft}
and for the magnetic quadrupole (M2) transitions: 
\end{flushleft}
\begin{equation}
A_{ji} = \frac{1.4910\times{10^{13}}}{{{\omega}_j}\lambda^5_{ji}} S \hspace*{1.0 cm} {\rm and} \hspace*{1.0 cm}
f_{ij} = \frac{2.236\times{10^{-3}}}{\lambda^3_{ji}\omega_i} S. \\
\end{equation}

In our calculations with the {\sc grasp} code the A- and f- values have been determined in both Babushkin and Coulomb gauges, which are  equivalent to the length and velocity forms in the more familiar non-relativistic nomenclature. However,  in Table 3 we present results in the length form alone, as  the velocity form is generally  considered to be comparatively less accurate.   Included in this table are the  transition energies ($\Delta$E$_{ij}$ in ${\rm \AA}$), radiative rates (A$_{ji}$ in s$^{-1}$), oscillator strengths ($f_{ij}$, dimensionless), and line strengths ($S$ in a.u.) for all 16,260 electric dipole (E1) transitions among the lowest 360 levels of W XL. The {\em indices} used to represent the lower and upper levels of a transition have already been defined in Table 2. Also, in calculating the  above parameters we have used the Breit and QED-corrected theoretical energies/wavelengths as listed in Table 2. However, only A- values are included in Table 3 for the 23,975 electric quadrupole (E2), 16,578  magnetic dipole (M1), and 23,501 magnetic quadrupole (M2) transitions. Corresponding results for f- or S- values can be  obtained by using Eqs. (1-5). 

A detailed comparison among a variety of f- value calculations with both {\sc grasp} and {\sc fac} has already been made in our earlier paper \cite{cjp}, but only for transitions from the lowest two levels  of the 4s$^2$4p$^5$ configuration to higher-lying ones. In Table D we provide a similar comparison for transitions from the lowest three levels of the 4s$^2$4p$^4$4d configuration. Included in this table are our results for f- values from the GRASP, FAC1 and FAC2 calculations. As noted earlier for energy levels in Table C, the effect of additional CI in FAC2 is small, because for all transitions (strong as well as weak) the f- values are comparable with those in FAC1.  Similarly, there is general agreement (within $\sim$ 20\%) between the GRASP and FAC results, particularly for those transitions for which the f- values are significant, i.e. $\ge$ 0.01. For some weak(er) transitions, such as 3--35/47/62, 4--52, and 5--37/40/47, the differences between the two sets of f- values are up to a factor of two. The f- values from FAC are sometimes higher (e.g. 3--47 and 3--59) and in other instances are lower (e.g. 3--35 and 3--62). Such discrepancies for some weak transitions are common among calculations with different codes, and examples may be found in some of our earlier papers, including that on  Mg-like ions \cite{fe15}.

Assessing the accuracy of atomic data is not a trivial task \cite{fst}. However, a  criterion commonly used to determine the reliability of f- (or A-) values is the agreement between their length and velocity forms. Although good agreement between the two forms is  desirable, it is not a necessary condition for accuracy. This is because different sets of configurations may lead to good agreement between the two forms, but entirely different results in magnitude, not only for the weaker inter-combination  transitions, but also the allowed ones which are comparatively  stable and larger in magnitude. Examples of this are provided in some of our earlier work, such as \cite{ah1}--\cite{kma}. Nevertheless, in Table D we list the velocity/length ratio of the f- values corresponding  to our GRASP calculations. For most transitions the two forms agree within $\sim$ 20\%, but discrepancies are larger (up to a factor of four) for some, such as 3--34/35/6, 4--34, and  5--32/37/40. Most of these are weak (f $<$ 0.01), but some are not, such as 5--49 and 5--53.  Based on the comparisons shown in Table D as well as in Table 3 of our earlier paper \cite{cjp}, we estimate the  accuracy of our radiative data to be better than 20\%, for a majority of transitions, particularly the stronger ones.

\begin{flushleft}
{\bf 4. Lifetimes}
\end{flushleft}

The lifetime $\tau$ of a level $j$ is defined as follows:

\begin{equation}
{\tau}_j = \frac{1}{{\sum_{i}^{}} A_{ji}}.
\end{equation}
In Table 2 we list lifetimes for all 360 levels from our calculations with the {\sc grasp} code, which  include A- values from all types of transitions, i.e. E1, E2, M1, and M2. Unfortunately, there are no measurements  or other theoretical results available with which to compare the lifetimes. However, we hope the data will be useful for future comparisons and further assessment of the accuracy of our radiative rates.

\begin{flushleft}
{\bf 5. Conclusions}
\end{flushleft}

In this work, energy levels and lifetimes obtained with the {\sc grasp} code have been reported for the lowest 360 levels of W XL, although calculations have been performed for up to 11,849 levels. Radiative rates, oscillator strengths, and line strengths have also been provided  for transitions among these levels. To assess the accuracy  of our  data, similar calculations have also been performed with the {\sc fac} code, as no other results are available in the literature.  Based on comparisons of calculations with the two independent codes, our energy levels are assessed to be accurate to $\sim$ 0.02 Ryd, whereas the accuracy for other atomic parameters is $\sim$ 20\%.

\section*{Acknowledgment}
 KMA  is thankful to  AWE Aldermaston for financial support.   



\newpage
\begin{flushleft}

\end{flushleft}


\newpage
\clearpage

\begin{table*}
\begin{flushleft}
{\bf Table A.} Mixing coefficients (MC) for the levels of the 4s$^2$4p$^5$,4s4p$^6$, and 4s$^2$4p$^4$4d configurations of W XL.  Numbers outside and inside a bracket correspond to MC and the level, respectively.  Level orderings correspond to those of Table 2.  \\
\end{flushleft}
\begin{tabular}{rllrrrrrrrrr} \hline
\\
Index  & Configuration           & Level              &  Mixing coefficients     \\
\\ \hline
\\
    1 &   4s$^2$4p$^5$           &  $^2$P$^o _{3/2 }$ &   0.99(  1)                                                                                           \\                                
                            
    2 &   4s$^2$4p$^5$           &  $^2$P$^o _{1/2 }$ &   0.99(  2)                             
                                                              \\
    3 &   4s$^2$4p$^4$($^3$P)4d  &  $^4$D$   _{3/2 }$ &  $-$0.61(  3)$+$0.43( 12)$-$0.29( 28)$+$0.44( 16)$-$0.26( 20)$-$0.23(  7)                             \\
    4 &   4s$^2$4p$^4$($^3$P)4d  &  $^4$P$   _{1/2 }$ &   0.37( 15)$-$0.68(  4)$-$0.24(  9)$+$0.45( 29)$-$0.32( 23)                                           \\
    5 &   4s$^2$4p$^4$($^3$P)4d  &  $^4$D$   _{5/2 }$ &  $-$0.35( 17) 0.66(  5)$-$0.26( 11)$+$0.26( 21)$+$0.40( 13)$-$0.33( 26)                               \\
    6 &   4s$^2$4p$^4$($^3$P)4d  &  $^4$F$   _{7/2 }$ &   0.56(  6)$-$0.31( 19)$+$0.52(  8)$+$0.52( 18)$-$0.20( 27)                                           \\
    7 &   4s$^2$4p$^4$($^1$S)4d  &  $^2$D$   _{3/2 }$ &   0.44( 60)$+$0.33( 12)$+$0.23( 28)$+$0.21( 22)$+$0.74(  7)                                           \\
    8 &   4s$^2$4p$^4$($^3$P)4d  &  $^2$F$   _{7/2 }$ &   0.65( 19)$+$0.51(  8)$+$0.25( 18)$+$0.47( 27)                                                       \\
    9 &   4s$^2$4p$^4$($^3$P)4d  &  $^2$P$   _{1/2 }$ &   0.25( 15)$+$0.27(  4)$-$0.60(  9)$+$0.36( 29)$+$0.48( 23)$-$0.38( 14)                               \\
   10 &   4s$^2$4p$^4$($^3$P)4d  &  $^4$F$   _{9/2 }$ &   0.83( 10)$+$0.55( 24)                 
                                                              \\
   11 &   4s$^2$4p$^4$($^3$P)4d  &  $^4$P$   _{5/2 }$ &   0.25( 17)$+$0.46( 11)$-$0.39( 25)$+$0.71( 66)                                                       \\
   12 &   4s$^2$4p$^4$($^3$P)4d  &  $^4$P$   _{3/2 }$ &   0.26( 60)$-$0.49( 12)$-$0.42( 28)$+$0.42( 22)$+$0.40( 16)$+$0.38( 20)                               \\
   13 &   4s$^2$4p$^4$($^1$D)4d  &  $^2$F$   _{5/2 }$ &  $-$0.35( 17)$-$0.39(  5)$+$0.26( 11)$-$0.22( 25)$+$0.48( 21)$+$0.43( 13)$+$0.30( 26)$-$0.32( 66)     \\
   14 &   4s4p$^6$               &  $^2$S$   _{1/2 }$ &   0.45(  4)$-$0.28(  9)$+$0.82( 14)     
                                                              \\
   15 &   4s$^2$4p$^4$($^3$P)4d  &  $^4$D$   _{1/2 }$ &  $-$0.89( 15)$-$0.31(  9)$+$0.27( 29)   
                                                              \\
   16 &   4s$^2$4p$^4$($^1$D)4d  &  $^2$D$   _{3/2 }$ &  $-$0.53( 60)$+$0.58(  3)$+$0.21( 12)$+$0.35( 22)$+$0.31( 16)$-$0.30( 20)                             \\
   17 &   4s$^2$4p$^4$($^3$P)4d  &  $^4$F$   _{5/2 }$ &   0.68( 17)$+$0.35( 25)$+$0.27( 21)$+$0.50( 13)$-$0.26( 26)                                           \\
   18 &   4s$^2$4p$^4$($^1$D)4d  &  $^2$G$   _{7/2 }$ &   0.37(  6)$+$0.40(  8)$-$0.77( 18)$+$0.29( 27)                                                       \\
   19 &   4s$^2$4p$^4$($^3$P)4d  &  $^4$D$   _{7/2 }$ &   0.62(  6)$+$0.65( 19)$-$0.33(  8)$-$0.27( 27)                                                       \\
   20 &   4s$^2$4p$^4$($^1$D)4d  &  $^2$P$   _{3/2 }$ &  $-$0.61( 12)$-$0.74( 20)               
                                                              \\
   21 &   4s$^2$4p$^4$($^3$P)4d  &  $^2$D$   _{5/2 }$ &   0.27( 17)$+$0.28(  5)$-$0.29( 11)$+$0.60( 21)$-$0.43( 13)$+$0.44( 26)                               \\
   22 &   4s$^2$4p$^4$($^3$P)4d  &  $^2$P$   _{3/2 }$ &  $-$0.32( 60)$-$0.37(  3) 0.69( 22)$-$0.50( 16)                                                       \\
   23 &   4s$^2$4p$^4$($^1$D)4d  &  $^2$S$   _{1/2 }$ &  $-$0.46(  4)$-$0.24(  9)$-$0.49( 29)$+$0.61( 23)$+$0.33( 14)                                         \\
   24 &   4s$^2$4p$^4$($^1$D)4d  &  $^2$G$   _{9/2 }$ &   0.55( 10)$-$0.83( 24)                 
                                                              \\
   25 &   4s$^2$4p$^4$($^3$P)4d  &  $^2$F$   _{5/2 }$ &   0.41(  5)$+$0.44( 11)$+$0.67( 25)$+$0.37( 26)                                                       \\
   26 &   4s$^2$4p$^4$($^1$D)4d  &  $^2$D$   _{5/2 }$ &   0.49( 11)$+$0.42( 21)$-$0.43( 13)$-$0.59( 26)                                                       \\
   27 &   4s$^2$4p$^4$($^1$D)4d  &  $^2$F$   _{7/2 }$ &  $-$0.34(  6)$+$0.20( 19)$+$0.45(  8)$-$0.25( 18)$-$0.76( 27)                                         \\
   28 &   4s$^2$4p$^4$($^3$P)4d  &  $^2$D$   _{3/2 }$ &   0.36( 60)$+$0.25(  3)$-$0.62( 28)$-$0.52( 16)$-$0.30( 20)                                           \\
   29 &   4s$^2$4p$^4$($^1$D)4d  &  $^2$P$   _{1/2 }$ &  $-$0.59(  9)$-$0.56( 29)$-$0.53( 23)   
                                                              \\
   60 &   4s$^2$4p$^4$($^3$P)4d  &  $^4$F$   _{3/2 }$ &  $-$0.45( 60)$-$0.24(  3)$-$0.48( 28)$-$0.36( 22)$+$0.58(  7)                                         \\
   66 &   4s$^2$4p$^4$($^1$S)4d  &  $^2$D$   _{5/2 }$ &  $-$0.35( 17)$-$0.37(  5)$-$0.36( 11)$+$0.43( 25)$+$0.23( 21)$+$0.60( 66)                             \\

 \\ \hline                                                                                            
\end{tabular}                                                                                   
\end{table*}

                               
\begin{flushleft}                                                                               
                               
{\small
                                                                               
}                                                                                               
                               
\end{flushleft} 

\newpage
\clearpage

\begin{table*}
\begin{flushleft}
{\bf Table B.}   Mixing coefficients (MC) for some levels of W XL.  Numbers outside and inside a bracket correspond to MC and the level, respectively. See Table 2 for level indices. \\
\end{flushleft}
 {\small
\begin{tabular}{rlllrrrrrrrr} \hline
Index  & Configuration           & Level              &  Mixing coefficients     \\
 \hline

   30  &  4s$^2$4p$^3$($^4$S)4d$^2$($^3$F)  &  $^6$F$^o _{3/2 }$  &     0.39( 30)$+$0.25( 80)$-$0.29(101)$-$0.25(154)$-$0.21(128)$-$0.38( 54)               \\                                       
       &                                    &                     &     $+$0.35(217)$-$0.24(125)$+$0.22( 46)                                                \\
   31  &  4s$^2$4p$^3$($^2$P)4d$^2$($^3$F)  &  $^4$D$^o _{1/2 }$  &    $-$0.48( 48)$-$0.22( 71)$-$0.24(113)$-$0.30(218)$-$0.58( 31)$-$0.30(272)             \\                                  
                               
   32  &  4s$^2$4p$^3$($^4$S)4d$^2$($^3$F)  &  $^6$F$^o _{5/2 }$  &    $-$0.39( 32)$-$0.26(117)$-$0.22(163)$-$0.34( 90)$-$0.24(153)$-$0.27( 95)             \\
       &                                    &                     &        $+$0.42(115)$-$0.20( 39)$+$0.26(230)$+$0.24(219)                                 \\
   33  &  4s$^2$4p$^3$($^2$D)4d$^2$($^3$F)  &  $^4$H$^o _{7/2 }$  &     0.30( 96)$+$0.23(227)$+$0.26(122)$-$0.35( 33)$-$0.24(250)$+$0.36(112)               \\
       &                                    &                     &        $-$0.24( 37)$+$0.33( 43)$+$0.34( 57)                                             \\
   34  &  4s$^2$4p$^3$($^4$S)4d$^2$($^3$P)  &  $^6$P$^o _{3/2 }$  &     0.38( 34)$+$0.24(156)$-$0.23(167)$-$0.23( 70)$+$0.26(116)$-$0.35(170)               \\
       &                                    &                     &        $+$0.25(237)$+$0.21(257)$+$0.32(290)                                             \\
   35  &  4s4p$^5$($^3$P)4d                 &  $^4$P$^o _{1/2 }$  &     0.53( 35)$+$0.30( 52)$-$0.39(162)$-$0.37(119)$+$0.34(245)                           \\                                  
                               
   36  &  4s$^2$4p$^3$($^2$P)4d$^2$($^3$F)  &  $^4$G$^o _{9/2 }$  &    $-$0.47( 97)$-$0.21(127)$+$0.33(235)$+$0.22(165)$-$0.53( 36)$+$0.30( 55)$-$0.20(254) \\  
   37  &  4s$^2$4p$^3$($^2$P)4d$^2$($^3$F)  &  $^4$F$^o _{7/2 }$  &    $-$0.30( 96)$-$0.30( 91)$-$0.22(112)$+$0.33( 37)$+$0.26(284)$-$0.23(264)$+$0.24( 57) \\
   38  &  4s$^2$4p$^3$($^2$D)4d$^2$($^3$F)  &  $^4$P$^o _{3/2 }$  &    $-$0.29( 89)$+$0.37( 58)$-$0.25( 30)$+$0.23(114)$-$0.22(108)$-$0.22( 38)             \\
       &                                    &                     &        $-$0.38(217)$+$0.23(293)$-$0.23(170)$-$0.22(257)                                 \\ 
   39  &  4s$^2$4p$^3$($^2$P)4d$^2$($^3$F)  &  $^4$D$^o _{5/2 }$  &    $-$0.20( 49)$+$0.28( 59)$+$0.20( 32)$-$0.26(117)$-$0.25( 90)$-$0.24( 45)             \\
       &                                    &                     &        $+$0.35( 39)$+$0.33(230)$+$0.20(330)                                             \\
   40  &  4s$^2$4p$^3$($^2$P)4d$^2$($^3$P)  &  $^4$D$^o _{5/2 }$  &    $-$0.50( 98)$+$0.23(251)$+$0.30(132)$-$0.41( 40)$+$0.44(238)                         \\                                  
                               
   41  &  4s$^2$4p$^3$($^4$S)4d$^2$($^3$P)  &  $^6$P$^o _{7/2 }$  &    $-$0.24( 79)$+$0.32( 41)$+$0.26( 84)$-$0.24(149)$+$0.39(264)$-$0.26(155)$+$0.27(186) \\
   42  &  4s$^2$4p$^3$($^4$S)4d$^2$($^3$F)  &  $^6$F$^o _{11/2}$  &     0.40( 42)$+$0.37(136)$-$0.21(103)$+$0.24(145)$-$0.33( 56)$+$0.49(239)$+$0.47(255)   \\
   43  &  4s$^2$4p$^3$($^2$P)4d$^2$($^3$F)  &  $^2$G$^o _{7/2 }$  &    $-$0.21( 68)$-$0.28( 96)$-$0.35( 33)$+$0.30( 37)$-$0.22(147)$+$0.36( 43)             \\                                  
                               
   44  &  4s$^2$4p$^4$($^1$D)4f             &  $^2$H$^o _{9/2 }$  &     0.26(176)$+$0.31( 73)$+$0.28( 44)$-$0.39(270)$-$0.35(179)$-$0.22(142)               \\
       &                                    &                     &        $-$0.29(121)$+$0.44(157)                                                         \\ 
   45  &  4s$^2$4p$^3$($^2$D)4d$^2$($^3$F)  &  $^4$F$^o _{5/2 }$  &    $-$0.20( 59)$+$0.27( 32)$-$0.20(274)$-$0.23(159)$-$0.28( 45)$+$0.27(251)             \\
       &                                    &                     &        $-$0.23(115)$+$0.26(230)$-$0.22( 40)$-$0.26(263)                                 \\
   46  &  4s$^2$4p$^3$($^2$P)4d$^2$($^1$D)  &  $^2$P$^o _{3/2 }$  &     0.20( 30)$-$0.25(100)$-$0.29( 80)$-$0.20(101)$-$0.21(146)$+$0.22(128)               \\
       &                                    &                     &        $+$0.22(126)$-$0.28( 54)$+$0.21(237)$+$0.30(318)$+$0.25(125)$-$0.30( 46)         \\ 
   47  &  4s$^2$4p$^4$($^1$D)4f             &  $^2$D$^o _{3/2 }$  &     0.24(171)$-$0.28(150)$+$0.26( 47)$+$0.28( 76)$-$0.30( 89)$+$0.27( 87)               \\
       &                                    &                     &        $-$0.25(114)$+$0.23(217)$-$0.24(293)                                             \\ 
   48  &  4s$^2$4p$^3$($^4$S)4d$^2$($^3$F)  &  $^6$F$^o _{1/2 }$  &     0.28( 61)$+$0.26( 88)$-$0.59( 35)$-$0.25( 48)$+$0.21( 71)                           \\
       &                                    &                     &        $-$0.23( 31)$+$0.27(272)                                                         \\         
   49  &  4s4p$^5$($^3$P)4d                 &  $^4$F$^o _{5/2 }$  &    $-$0.35(151)$+$0.22( 72)$+$0.28(187)$-$0.42( 49)$+$0.24( 59)                         \\
       &                                    &                     &        $+$0.25(  0)$-$0.26( 85)$-$0.21(152)$-$0.33(330)                                 \\         
   50  &  4s$^2$4p$^3$($^2$D)4d$^2$($^1$D)  &  $^2$F$^o _{5/2 }$  &    $-$0.21(168)$-$0.23( 53)$-$0.23( 67)$-$0.27(163)$-$0.29( 50)                         \\
       &                                    &                     &        $-$0.21(324)$+$0.35(219)                                                         \\         
   51  &  4s4p$^5$($^3$P)4d                 &  $^4$F$^o _{7/2 }$  &     0.38( 68)$+$0.56( 51)$-$0.25( 75)$+$0.31( 79)                                       \\                                  
                               
   52  &  4s$^2$4p$^3$($^4$S)4d$^2$($^3$P)  &  $^4$P$^o _{1/2 }$  &     0.45( 35)$+$0.32(189)$-$0.35( 52)$-$0.21(106)$-$0.30( 62)                           \\
       &                                    &                     &        $+$0.24(119)$+$0.28(306)                                                         \\         
   53  &  4s$^2$4p$^4$($^1$S)4f             &  $^2$F$^o _{5/2 }$  &    $-$0.20(151)$-$0.28( 53)$+$0.23( 77)$-$0.22(159)$+$0.31( 85)                         \\
       &                                    &                     &        $+$0.26(152)$-$0.21(238)$-$0.21(263)$+$0.39(330)                                 \\         
   54  &  4s$^2$4p$^3$($^2$P)4d$^2$($^3$F)  &  $^4$F$^o _{3/2 }$  &    $-$0.22(150)$-$0.34( 89)$+$0.33( 58)$-$0.35( 34)$-$0.30(167)                         \\
       &                                    &                     &        $-$0.29(237)$+$0.32(257)                                                         \\         
   55  &  4s$^2$4p$^3$($^2$P)4d$^2$($^3$F)  &  $^4$F$^o _{9/2 }$  &     0.40( 65)$+$0.24(184)$+$0.63( 74)$-$0.27( 55)$-$0.22(254)                           \\                                  
                               
   56  &  4s$^2$4p$^3$($^2$D)4d$^2$($^1$G)  &  $^2$I$^o _{11/2}$  &     0.39( 69)$+$0.27(175)$+$0.37( 42)$-$0.31(136)$+$0.20(145)                           \\
       &                                    &                     &        $+$0.30( 56)$+$0.43(239)$-$0.42(255)                                             \\         
   57  &  4s$^2$4p$^3$($^2$P)4d$^2$($^1$D)  &  $^2$F$^o _{7/2 }$  &     0.24(325)$+$0.26( 75)$+$0.28( 79)$-$0.21(227)$+$0.27( 41)                           \\
       &                                    &                     &        $+$0.22(122)$-$0.23(250)$+$0.23(147)$+$0.31(264)$+$0.32( 57)                     \\         
   58  &  4s4p$^5$($^3$P)4d                 &  $^4$P$^o _{3/2 }$  &     0.25(171)$-$0.25( 63)$+$0.29( 47)$-$0.56( 58)$+$0.33( 87)                           \\                                  
                               
   59  &  4s4p$^5$($^3$P)4d                 &  $^4$D$^o _{5/2 }$  &     0.24(168)$+$0.27( 53)$+$0.48( 59)$-$0.26( 67)$+$0.25( 77)$-$0.37( 78)               \\         
   61  &  4s$^2$4p$^4$($^3$P)4f             &  $^4$D$^o _{1/2 }$  &     0.54( 61)$+$0.40(148)$+$0.34( 88)$+$0.26(189)$-$0.28(272)                           \\                                  
                               
   62  &  4s$^2$4p$^3$($^2$D)4d$^2$($^3$P)  &  $^4$D$^o _{1/2 }$  &    $-$0.22( 86)$-$0.29(130)$-$0.28( 71)$-$0.41( 62)$-$0.32(162)                         \\
       &                                    &                     &        $+$0.40(306)$+$0.28(245)$-$0.32(272)                                             \\         
   63  &  4s$^2$4p$^4$($^3$P)4f             &  $^2$D$^o _{3/2 }$  &     0.36(150)$-$0.44( 63)$+0.43(161)$-$0.21( 83)$+0.34(188)                             \\                                  
                               
   64  &  4s$^2$4p$^4$($^3$P)4f             &  $^2$F$^o _{7/2 }$  &    $-$0.27( 68)$+$0.22(185)$+$0.38( 64)$+$0.28(164)$+$0.26(234)                         \\
       &                                    &                     &        $-$0.21(250)$-$0.21(147)$-$0.25(284)$+$0.24( 57)                                 \\         
   65  &  4s$^2$4p$^4$($^3$P)4f             &  $^4$F$^o _{9/2 }$  &    $-$0.27(176)$+$0.41( 65)$+$0.29(184)$-$0.27( 97)$+$0.24(127)                         \\
       &                                    &                     &        $-$0.27(235)$-$0.29( 92)$+$0.37( 55)$+$0.34(254)                                 \\         

\hline                                                                                                                                  
\end{tabular}      
}                                                                             
\end{table*}    


\newpage
\clearpage

\begin{table*}
\begin{flushleft}
{\bf Table C.}  Comparison of excitation energies (in Ryd) for some levels of W XL.   Level orderings correspond to those of Table 2.  \\
\end{flushleft}
\begin{tabular}{rllrrrrrrrrrr} \hline
Index  & Configuration                       & Level              & GRASP    & FAC1    & FAC2     \\
 \hline
   30 &   4s$^2$4p$^3$($^4$S)4d$^2$($^3$F)   &  $^6$F$^o _{3/2 }$ &  22.5740 & 22.5805 & 22.5916  \\ 
   31 &   4s$^2$4p$^3$($^2$P)4d$^2$($^3$F)   &  $^4$D$^o _{1/2 }$ &  22.7106 & 22.7166 & 22.7278  \\ 
   32 &   4s$^2$4p$^3$($^4$S)4d$^2$($^3$F)   &  $^6$F$^o _{5/2 }$ &  22.7488 & 22.7545 & 22.7658  \\ 
   33 &   4s$^2$4p$^3$($^2$D)4d$^2$($^3$F)   &  $^4$H$^o _{7/2 }$ &  22.8911 & 22.8951 & 22.9063  \\ 
   34 &   4s$^2$4p$^3$($^4$S)4d$^2$($^3$P)   &  $^6$P$^o _{3/2 }$ &  23.3766 & 23.3797 & 23.3891  \\ 
   35 &   4s4p$^5$($^3$P)4d                  &  $^4$P$^o _{1/2 }$ &  24.0349 & 24.0451 & 24.0555  \\ 
   36 &   4s$^2$4p$^3$($^2$P)4d$^2$($^3$F)   &  $^4$G$^o _{9/2 }$ &  24.0548 & 24.0666 & 24.0781  \\ 
   37 &   4s$^2$4p$^3$($^2$P)4d$^2$($^3$F)   &  $^4$F$^o _{7/2 }$ &  24.0662 & 24.0779 & 24.0890  \\ 
   38 &   4s$^2$4p$^3$($^2$D)4d$^2$($^3$F)   &  $^4$P$^o _{3/2 }$ &  24.0928 & 24.1036 & 24.1143  \\ 
   39 &   4s$^2$4p$^3$($^2$P)4d$^2$($^3$F)   &  $^4$D$^o _{5/2 }$ &  24.2275 & 24.2371 & 24.2480  \\ 
   40 &   4s$^2$4p$^3$($^2$P)4d$^2$($^3$P)   &  $^4$D$^o _{5/2 }$ &  24.3977 & 24.4083 & 24.4189  \\ 
   41 &   4s$^2$4p$^3$($^4$S)4d$^2$($^3$P)   &  $^6$P$^o _{7/2 }$ &  24.4336 & 24.4426 & 24.4532  \\ 
   42 &   4s$^2$4p$^3$($^4$S)4d$^2$($^3$F)   &  $^6$F$^o _{11/2}$ &  24.5388 & 24.5484 & 24.5595  \\ 
   43 &   4s$^2$4p$^3$($^2$P)4d$^2$($^3$F)   &  $^2$G$^o _{7/2 }$ &  24.9901 & 24.9957 & 25.0069  \\ 
   44 &   4s$^2$4p$^4$($^1$D)4f              &  $^2$H$^o _{9/2 }$ &  24.9948 & 25.0028 & 25.0136  \\ 
   45 &   4s$^2$4p$^3$($^2$D)4d$^2$($^3$F)   &  $^4$F$^o _{5/2 }$ &  25.0279 & 25.0330 & 25.0439  \\ 
   46 &   4s$^2$4p$^3$($^2$P)4d$^2$($^1$D)   &  $^2$P$^o _{3/2 }$ &  25.2438 & 25.2483 & 25.2591  \\ 
   47 &   4s$^2$4p$^4$($^1$D)4f              &  $^2$D$^o _{3/2 }$ &  25.2599 & 25.2588 & 25.2689  \\ 
   48 &   4s$^2$4p$^3$($^4$S)4d$^2$($^3$F)   &  $^6$F$^o _{1/2 }$ &  25.2688 & 25.2699 & 25.2803  \\ 
   49 &   4s4p$^5$($^3$P)4d                  &  $^4$F$^o _{5/2 }$ &  25.2857 & 25.2857 & 25.2958  \\ 
   50 &   4s$^2$4p$^3$($^2$D)4d$^2$($^1$D)   &  $^2$F$^o _{5/2 }$ &  25.4722 & 25.4819 & 25.4926  \\ 
   51 &   4s4p$^5$($^3$P)4d                  &  $^4$F$^o _{7/2 }$ &  25.4883 & 25.4873 & 25.4972  \\ 
   52 &   4s$^2$4p$^3$($^4$S)4d$^2$($^3$P)   &  $^4$P$^o _{1/2 }$ &  25.5114 & 25.5113 & 25.5215  \\ 
   53 &   4s$^2$4p$^4$($^1$S)4f              &  $^2$F$^o _{5/2 }$ &  25.6214 & 25.6295 & 25.6399  \\ 
   54 &   4s$^2$4p$^3$($^2$P)4d$^2$($^3$F)   &  $^4$F$^o _{3/2 }$ &  25.7739 & 25.7706 & 25.7807  \\ 
   55 &   4s$^2$4p$^3$($^2$P)4d$^2$($^3$F)   &  $^4$F$^o _{9/2 }$ &  25.8007 & 25.8092 & 25.8194  \\ 
   56 &   4s$^2$4p$^3$($^2$D)4d$^2$($^1$G)   &  $^2$I$^o _{11/2}$ &  25.8381 & 25.8518 & 25.8626  \\ 
   57 &   4s$^2$4p$^3$($^2$P)4d$^2$($^1$D)   &  $^2$F$^o _{7/2 }$ &  25.9381 & 25.9510 & 25.9616  \\ 
   58 &   4s4p$^5$($^3$P)4d                  &  $^4$P$^o _{3/2 }$ &  26.0483 & 26.0447 & 26.0545  \\ 
   59 &   4s4p$^5$($^3$P)4d                  &  $^4$D$^o _{5/2 }$ &  26.3546 & 26.3458 & 26.3550  \\ 
   61 &   4s$^2$4p$^4$($^3$P)4f              &  $^4$D$^o _{1/2 }$ &  26.4363 & 26.4309 & 26.4398  \\ 
   62 &   4s$^2$4p$^3$($^2$D)4d$^2$($^3$P)   &  $^4$D$^o _{1/2 }$ &  26.7300 & 26.7355 & 26.7458  \\ 
   63 &   4s$^2$4p$^4$($^3$P)4f              &  $^2$D$^o _{3/2 }$ &  26.7611 & 26.7650 & 26.7740  \\ 
   64 &   4s$^2$4p$^4$($^3$P)4f              &  $^2$F$^o _{7/2 }$ &  26.7815 & 26.7908 & 26.8009  \\ 
   65 &   4s$^2$4p$^4$($^3$P)4f              &  $^4$F$^o _{9/2 }$ &  26.8418 & 26.8461 & 26.8566  \\ 
 \hline                                                                                         
         
\end{tabular}



\begin{flushleft}                                                                               
                               
{\small
GRASP: present calculations from the {\sc grasp} code with 4215 levels\\ 
FAC1: present calculations from the {\sc fac} code with 4215 levels  \\  
FAC2: present calculations from the {\sc fac} code with 11,849 levels \\

}                                                                                               
                               
\end{flushleft} 

\end{table*}


\newpage
\clearpage

\begin{table*}
\begin{flushleft}
{\bf Table D.}  Comparison of oscillator strengths (f- values) for some transitions of W XL. See Table 2 for level indices.  $a{\pm}b \equiv a{\times}$10$^{{\pm}b}$.  \\
\end{flushleft}

\begin{tabular}{rrrrrrrrrrrr} \hline
     I &   J    & GRASP      &       FAC1 &       FAC2 & Ratio   &     I  &   J     & GRASP      &       FAC1 &       FAC2 & Ratio   \\
 \hline                                                                         
     3 &   30   & 3.9112-3  &  3.926-3  &  3.927-3  & 1.1-0   &     4 &   52   & 7.6985-3  &  5.016-3  &  5.018-3  & 7.5-1  \\
     3 &   31   & 1.4208-3  &  1.396-3  &  1.396-3  & 1.1-0   &     4 &   54   & 1.0811-1  &  1.066-1  &  1.068-1  & 1.1-0  \\
     3 &   34   & 1.1878-5  &  1.145-5  &  1.135-5  & 1.5-0   &     4 &   58   & 4.5795-2  &  4.783-2  &  4.791-2  & 1.1-0  \\
     3 &   35   & 9.0205-4  &  5.882-4  &  5.849-4  & 1.9-0   &     4 &   61   & 1.4751-2  &  1.145-2  &  1.155-2  & 1.9-0  \\
     3 &   38   & 1.0258-2  &  9.486-3  &  9.458-3  & 1.3-0   &     4 &   62   & 1.8444-2  &  1.725-2  &  1.722-2  & 1.0-0  \\
     3 &   39   & 2.7366-2  &  2.587-2  &  2.575-2  & 1.2-0   &     4 &   63   & 3.4014-3  &  2.880-3  &  2.923-3  & 9.2-1  \\
     3 &   40   & 3.7872-4  &  3.306-4  &  3.293-4  & 9.1-1   &     5 &   30   & 2.0459-3  &  2.051-3  &  2.054-3  & 1.5-0  \\
     3 &   45   & 1.9742-1  &  2.048-1  &  2.048-1  & 8.3-1   &     5 &   32   & 1.0676-4  &  1.194-4  &  1.204-4  & 6.6-1  \\
     3 &   46   & 7.8150-2  &  7.612-2  &  7.557-2  & 9.8-1   &     5 &   33   & 2.5187-3  &  2.633-3  &  2.630-3  & 1.4-0  \\
     3 &   47   & 5.9716-3  &  9.911-3  &  1.048-2  & 6.7-1   &     5 &   34   & 4.0929-3  &  3.965-3  &  3.961-3  & 1.3-0  \\
     3 &   48   & 2.7463-2  &  3.123-2  &  3.126-2  & 7.4-1   &     5 &   37   & 8.3477-5  &  6.747-5  &  6.851-5  & 4.3-0  \\
     3 &   49   & 2.6154-3  &  2.169-3  &  2.066-3  & 8.9-2   &     5 &   38   & 1.2041-3  &  9.124-4  &  9.082-4  & 9.8-1  \\
     3 &   50   & 2.5621-2  &  2.616-2  &  2.624-2  & 9.6-1   &     5 &   39   & 2.6589-3  &  2.290-3  &  2.285-3  & 1.6-0  \\
     3 &   52   & 3.9984-2  &  3.936-2  &  3.937-2  & 8.3-1   &     5 &   40   & 6.4753-5  &  5.028-5  &  5.096-5  & 2.9-0  \\
     3 &   53   & 1.2948-2  &  1.173-2  &  1.174-2  & 1.2-0   &     5 &   41   & 1.2505-3  &  1.130-3  &  1.122-3  & 1.7-0  \\
     3 &   54   & 6.0372-2  &  6.791-2  &  6.780-2  & 6.8-1   &     5 &   43   & 1.9014-1  &  1.950-1  &  1.947-1  & 8.5-1  \\
     3 &   58   & 4.9802-2  &  4.512-2  &  4.522-2  & 1.2-0   &     5 &   45   & 4.4200-2  &  4.139-2  &  4.131-2  & 1.0-0  \\
     3 &   59   & 2.5836-4  &  4.468-4  &  4.464-4  & 9.4-1   &     5 &   46   & 8.9965-3  &  9.916-3  &  9.895-3  & 1.0-0  \\
     3 &   61   & 4.6960-3  &  5.704-3  &  5.667-3  & 3.1-1   &     5 &   47   & 1.3296-4  &  9.793-5  &  1.170-4  & 1.9-0  \\
     3 &   62   & 3.6121-5  &  1.570-5  &  1.588-5  & 5.8-0   &     5 &   49   & 1.8391-2  &  2.491-2  &  2.509-2  & 6.0-1  \\
     3 &   63   & 4.0475-3  &  3.542-3  &  3.575-3  & 9.3-1   &     5 &   50   & 2.2852-2  &  2.426-2  &  2.421-2  & 6.9-1  \\
     4 &   30   & 8.3655-3  &  8.354-3  &  8.344-3  & 1.3-0   &     5 &   51   & 1.6155-2  &  1.480-2  &  1.495-2  & 1.5-0  \\
     4 &   31   & 1.8952-2  &  1.884-2  &  1.882-2  & 1.2-0   &     5 &   53   & 1.5529-2  &  1.468-2  &  1.458-2  & 5.6-1  \\
     4 &   34   & 2.1947-4  &  1.880-4  &  1.847-4  & 3.2-2   &     5 &   54   & 6.2702-2  &  6.808-2  &  6.809-2  & 9.3-1  \\
     4 &   35   & 3.2561-3  &  2.644-3  &  2.639-3  & 1.7-0   &     5 &   57   & 4.1634-4  &  3.568-4  &  3.562-4  & 6.0-1  \\
     4 &   38   & 4.4388-3  &  3.927-3  &  3.907-3  & 1.7-0   &     5 &   58   & 4.8795-2  &  4.782-2  &  4.779-2  & 6.4-1  \\
     4 &   46   & 1.9806-1  &  2.110-1  &  2.143-1  & 8.3-1   &     5 &   59   & 6.6341-2  &  6.831-2  &  6.829-2  & 9.1-1  \\
     4 &   47   & 1.2180-1  &  1.158-1  &  1.124-1  & 7.6-1   &     5 &   63   & 3.2631-3  &  3.325-3  &  3.354-3  & 8.2-1  \\
     4 &   48   & 6.0750-2  &  6.826-2  &  6.827-2  & 8.3-1   &     5 &   64   & 9.6953-3  &  9.421-3  &  9.327-3  & 9.1-1  \\
 \hline                                                                                               
\end{tabular}                                                                                   
                               
\newpage                                                                                                
                               
                               
\begin{flushleft}                                                                               
                               
{\small
GRASP: present calculations from the {\sc grasp} code with 4215 levels\\ 
FAC1: present calculations from the {\sc fac} code with 4215 levels  \\  
FAC2: present calculations from the {\sc fac} code with 11,849 levels \\    
Ratio: ratio of velocity and length forms of f- values from the {\sc grasp} code \\                                                                                 
}                                                                                               
                               
\end{flushleft} 

\end{table*}


\newpage
\clearpage

\begin{flushleft}
{\bf Explanation of Tables} \\  \vspace{0.2 cm}
\end{flushleft}

\begin{flushleft}
Table 1. Configurations and  levels of W XL.  \\ \vspace{0.2 cm}
\begin{tabular}{ll}
Index            &  Configuration Index \\
Configuration    & The configuration included in the calculation \\
No. of Levels             & The number of levels the configuration generates. Odd parity levels are designated with a superscript ``o" \\
Energy  Range         & Energy range of the levels in Ryd unit \\
GRASPa            &     earlier calculations \cite{cjp} from the {\sc grasp} code with 4128 levels \\ 
GRASPb           &      present calculations from the {\sc grasp} code with 4215 levels\\										
RELAC              &      Fournier \cite{kbf} \\ 
Y                          &     If a configuration is included under a calculation \\ 
\end{tabular}
\end{flushleft}

\begin{flushleft}
Table 2. Energies (Ryd) for the lowest 360 levels of W XL and their lifetimes ($\tau$, s).  \\ \vspace{0.2 cm}
\begin{tabular}{ll}
Index            & Level Index \\
Configuration    & The configuration to which the level belongs \\
Level             & The $LSJ$ designation of the level \\
Energy           & Present energies from the {\sc grasp} code  with 4215 level calculations \\
$\tau$ (s)       & Lifetime of the level in s \\
\end{tabular}
\end{flushleft}

\begin{flushleft}
Table 3. Transition wavelengths ($\lambda_{ij}$ in $\rm \AA$), radiative rates (A$_{ji}$ in s$^{-1}$),
 oscillator strengths (f$_{ij}$, dimensionless), and line strengths (S, in atomic units) for electric dipole (E1), and 
A$_{ji}$ for electric quadrupole (E2), magnetic dipole (M1), and magnetic quadrupole (M2) transitions of W XL. \\ \vspace{0.2 cm}
\begin{tabular}{ll}
$i$ and $j$         & The lower ($i$) and upper ($j$) levels of a transition as defined in Table 2.\\
$\lambda_{ij}$      & Transition wavelength (in ${\rm \AA}$) \\
A$^{E1}_{ji}$       & Radiative transition probability (in s$^{-1}$) for the E1 transitions \\
f$^{E1}_{ij}$       & Absorption oscillator strength (dimensionless) for the E1 transitions \\
S$^{E1}$            & Line strength in atomic unit (a.u.), 1 a.u. = 6.460$\times$10$^{-36}$ cm$^2$ esu$^2$ for the E1 transitions \\
A$^{E2}_{ji}$       & Radiative transition probability (in s$^{-1}$) for the E2 transitions \\
A$^{M1}_{ji}$       & Radiative transition probability (in s$^{-1}$) for the M1 transitions \\
A$^{M2}_{ji}$       & Radiative transition probability (in s$^{-1}$) for the M2 transitions \\
$a{\pm}b$ &  $\equiv a\times{10^{{\pm}b}}$ \\
\end{tabular}
\end{flushleft}


\newpage
\clearpage

\begin{flushleft}
Table 1. Configurations and  levels of W XL. 
\end{flushleft}
\begin{tabular}{rllcrrrrrrrrr} \hline
Index  & Configuration      & No. of Levels & Energy Range (Ryd)&  GRASPa & GRASPb   &  RELAC      \\
 \hline
  1  &  4s$^2$4p$^5$ 		        &     2$^o$  & 0--7	& Y  & Y & Y \\
  2  &  4s$^2$4p$^4$4d 		        &    28      & 11--27	& Y  & Y & Y \\
  3  &  4s$^2$4p$^4$4f 		        &    30$^o$  & 25--42	& Y  & Y & Y \\
  4  &  4s4p$^6$		        &     1      & 15	& Y  & Y & Y \\
  5  &  4p$^6$4d		        &     2      & 43--45	& Y  & Y &   \\
  6  &  4p$^6$4f		        &     2$^o$  & 58--59	& Y  & Y &   \\
  7  &  4s4p$^5$4d		        &    23$^o$  & 25--36	& Y  & Y & Y \\
  8  &  4s4p$^5$4f		        &    24      & 41--50	& Y  & Y & Y \\
  9  &  4s$^2$4p$^3$4d$^2$		&   141$^o$  & 22--42	& Y  & Y &   \\
 10  &  4s$^2$4p$^3$4f$^2$		&   221$^o$  & 52--71	& Y  & Y &   \\
 11  &  4s$^2$4p$^3$4d4f		&   363      & 37--58	& Y  & Y &   \\
 12  &  4s$^2$4p$^2$4d$^3$		&   261      & 34--57	& Y  & Y &   \\
 13  &  4s$^2$4p4d$^4$		        &   180$^o$  & 54--73	& Y  & Y &   \\
 14  &  4s$^2$4p$^2$4d$^2$4f		&  1140$^o$  & 49--73	& Y  & Y &   \\
 15  &  4s4p$^3$4d$^3$		        &   678$^o$  & 47--73	& Y  & Y &   \\
 16  &  4p$^5$4d$^2$		        &    45$^o$  & 53--64	& Y  & Y &   \\
 17  &  3d$^9$4s$^2$4p$^5$4d	        &    96$^o$  & 131--147 & Y  & Y & Y \\
 18  &  3d$^9$4s$^2$4p$^5$4f	        &   113      & 147--161 & Y  & Y & Y \\
 19  &  3d$^9$4s$^2$4p$^6$	        &     2      & 120--126 & Y  & Y & Y \\
 20  &  3p$^5$3d$^{10}$4s$^2$4p$^6$     &     2$^o$  & 155--178 & Y  & Y & Y \\
 21  &  3p$^5$3d$^{10}$4s$^2$4p$^5$4d   &    65      & 165--198 & Y  & Y & Y \\
 22  &  3p$^5$3d$^{10}$4s$^2$4p$^5$4f   &    36$^o$  & 180--212 & Y  & Y & Y \\
 23  &  4s4p$^5$5s		        &     7$^o$  & 65--74	& Y  & Y &   \\
 24  &  4s4p$^5$5p		        &    18      & 68--80	& Y  & Y &   \\
 25  &  4s4p$^5$5d		        &    23$^o$  & 77--86	& Y  & Y &   \\
 26  &  4s4p$^5$5f		        &    24      & 84--92	& Y  & Y &   \\
 27  &  4s4p$^5$5g		        &    24$^o$  & 87--96	& Y  & Y &   \\
 28  &  4p$^6$5s		        &     1      & 82	& Y  & Y &   \\
 29  &  4p$^6$5p		        &     2$^o$  & 86--89	& Y  & Y &   \\
 30  &  4p$^6$5d		        &     2      & 94--95	& Y  & Y &   \\
 31  &  4p$^6$5f		        &     2$^o$  & 100--101 & Y  & Y &   \\
 32  &  4p$^6$5g		        &     2      & 104--105 & Y  & Y &   \\
 33  &  4s$^2$4p$^4$5s		        &     8      & 50--65	& Y  & Y & Y \\
 34  &  4s$^2$4p$^4$5p		        &    21$^o$  & 54--72	& Y  & Y & Y \\
 35  &  4s$^2$4p$^4$5d		        &    28      & 62--78	& Y  & Y & Y \\
 36  &  4s$^2$4p$^4$5f		        &    30$^o$  & 69--84	& Y  & Y & Y \\
 37  &  4s$^2$4p$^4$5g		        &    30      & 73--88	& Y  & Y &   \\
 38  &  3d$^9$4s$^2$4p$^5$5s	        &    23$^o$  & 172--185 & Y  & Y & Y \\
 39  &  3d$^9$4s$^2$4p$^5$5p	        &    65      & 175--191 & Y  & Y & Y \\
 40  &  3d$^9$4s$^2$4p$^5$5d	        &    96$^o$  & 183--197 & Y  & Y & Y \\
 41  &  3d$^9$4s$^2$4p$^5$5f	        &   113      & 190--203 & Y  & Y & Y \\
 42  &  3d$^9$4s$^2$4p$^5$5g	        &   119$^o$  & 194--207 & Y  & Y &   \\
 43  &  4s$^2$4p$^4$6s 		        &    8       & 81--96   &    & Y &   \\
 44  &  4s$^2$4p$^4$6p 		        &   21$^o$   & 81--99   &    & Y &   \\
 45  &  4s$^2$4p$^4$6d 		        &   28       & 87--102  &    & Y &   \\
 46  &  4s$^2$4p$^4$6f 		        &   30$^o$   & 91--106  &    & Y &   \\
 \hline  											      
\end{tabular}   								   					       
			      							   					       
\begin{flushleft}													       
{\small
GRASPa: earlier calculations \cite{cjp} from the {\sc grasp} code with 4128 levels \\ 
GRASPb: present calculations from the {\sc grasp} code with 4215 levels\\										
RELAC: Fournier \cite{kbf} \\ 
Y: configuration included under a calculation \\ 										       
															       
}															       
\end{flushleft} 

\clearpage
\newpage

\begin{flushleft}
Table 2. Energies (Ryd) for the lowest 360 levels of W XL and their lifetimes ($\tau$, s). $a{\pm}b \equiv a{\times}$10$^{{\pm}b}$.
\end{flushleft}
\begin{tabular}{rllrrrrrrrrrr} \hline
Index  & Configuration                       & Level              & Energy  & $\tau$  (s)   \\
 \hline
    1 &   4s$^2$4p$^5$  		     &  $^2$P$^o _{3/2 }$ &   0.0000 &   ........  \\ 
    2 &   4s$^2$4p$^5$  		     &  $^2$P$^o _{1/2 }$ &   6.7987 &   1.314-07  \\ 
    3 &   4s$^2$4p$^4$($^3$P)4d 	     &  $^4$D$   _{3/2 }$ &  11.2603 &   3.216-10  \\ 
    4 &   4s$^2$4p$^4$($^3$P)4d 	     &  $^4$P$   _{1/2 }$ &  11.4314 &   7.153-11  \\ 
    5 &   4s$^2$4p$^4$($^3$P)4d 	     &  $^4$D$   _{5/2 }$ &  11.4275 &   6.342-10  \\ 
    6 &   4s$^2$4p$^4$($^3$P)4d 	     &  $^4$F$   _{7/2 }$ &  11.6140 &   4.757-03  \\ 
    7 &   4s$^2$4p$^4$($^1$S)4d 	     &  $^2$D$   _{3/2 }$ &  12.1537 &   2.679-09  \\ 
    8 &   4s$^2$4p$^4$($^3$P)4d 	     &  $^2$F$   _{7/2 }$ &  12.7500 &   2.237-05  \\ 
    9 &   4s$^2$4p$^4$($^3$P)4d 	     &  $^2$P$   _{1/2 }$ &  12.8178 &   1.040-10  \\ 
   10 &   4s$^2$4p$^4$($^3$P)4d 	     &  $^4$F$   _{9/2 }$ &  12.8147 &   3.704-05  \\ 
   11 &   4s$^2$4p$^4$($^3$P)4d 	     &  $^4$P$   _{5/2 }$ &  13.7448 &   3.062-10  \\ 
   12 &   4s$^2$4p$^4$($^3$P)4d 	     &  $^4$P$   _{3/2 }$ &  14.0956 &   2.361-12  \\ 
   13 &   4s$^2$4p$^4$($^1$D)4d 	     &  $^2$F$   _{5/2 }$ &  14.3278 &   1.752-12  \\ 
   14 &   4s4p$^6$			     &  $^2$S$   _{1/2 }$ &  15.3699 &   1.656-12  \\ 
   15 &   4s$^2$4p$^4$($^3$P)4d 	     &  $^4$D$   _{1/2 }$ &  17.6737 &   1.220-10  \\ 
   16 &   4s$^2$4p$^4$($^1$D)4d 	     &  $^2$D$   _{3/2 }$ &  18.0340 &   5.114-10  \\ 
   17 &   4s$^2$4p$^4$($^3$P)4d 	     &  $^4$F$   _{5/2 }$ &  18.3401 &   2.068-11  \\ 
   18 &   4s$^2$4p$^4$($^1$D)4d 	     &  $^2$G$   _{7/2 }$ &  18.4099 &   2.604-07  \\ 
   19 &   4s$^2$4p$^4$($^3$P)4d 	     &  $^4$D$   _{7/2 }$ &  19.3377 &   1.356-07  \\ 
   20 &   4s$^2$4p$^4$($^1$D)4d 	     &  $^2$P$   _{3/2 }$ &  19.6012 &   1.315-12  \\ 
   21 &   4s$^2$4p$^4$($^3$P)4d 	     &  $^2$D$   _{5/2 }$ &  19.7541 &   5.212-13  \\ 
   22 &   4s$^2$4p$^4$($^3$P)4d 	     &  $^2$P$   _{3/2 }$ &  19.7592 &   1.001-12  \\ 
   23 &   4s$^2$4p$^4$($^1$D)4d 	     &  $^2$S$   _{1/2 }$ &  19.8287 &   3.712-13  \\ 
   24 &   4s$^2$4p$^4$($^1$D)4d 	     &  $^2$G$   _{9/2 }$ &  19.7982 &   2.445-07  \\ 
   25 &   4s$^2$4p$^4$($^3$P)4d 	     &  $^2$F$   _{5/2 }$ &  19.8152 &   1.731-11  \\ 
   26 &   4s$^2$4p$^4$($^1$D)4d 	     &  $^2$D$   _{5/2 }$ &  20.0716 &   1.664-11  \\ 
   27 &   4s$^2$4p$^4$($^1$D)4d 	     &  $^2$F$   _{7/2 }$ &  20.3657 &   2.046-07  \\ 
   28 &   4s$^2$4p$^4$($^3$P)4d 	     &  $^2$D$   _{3/2 }$ &  21.3920 &   1.224-12  \\ 
   29 &   4s$^2$4p$^4$($^1$D)4d 	     &  $^2$P$   _{1/2 }$ &  21.8225 &   7.443-13  \\ 
   30 &   4s$^2$4p$^3$($^4$S)4d$^2$($^3$F)   &  $^6$F$^o _{3/2 }$ &  22.5740 &   7.837-11  \\ 
   31 &   4s$^2$4p$^3$($^2$P)4d$^2$($^3$F)   &  $^4$D$^o _{1/2 }$ &  22.7106 &   3.897-11  \\ 
   32 &   4s$^2$4p$^3$($^4$S)4d$^2$($^3$F)   &  $^6$F$^o _{5/2 }$ &  22.7488 &   8.513-11  \\ 
   33 &   4s$^2$4p$^3$($^2$D)4d$^2$($^3$F)   &  $^4$H$^o _{7/2 }$ &  22.8911 &   2.090-10  \\ 
   34 &   4s$^2$4p$^3$($^4$S)4d$^2$($^3$P)   &  $^6$P$^o _{3/2 }$ &  23.3766 &   1.134-10  \\ 
   35 &   4s4p$^5$($^3$P)4d		     &  $^4$P$^o _{1/2 }$ &  24.0349 &   6.245-11  \\ 
   36 &   4s$^2$4p$^3$($^2$P)4d$^2$($^3$F)   &  $^4$G$^o _{9/2 }$ &  24.0548 &   1.186-10  \\ 
   37 &   4s$^2$4p$^3$($^2$P)4d$^2$($^3$F)   &  $^4$F$^o _{7/2 }$ &  24.0662 &   6.165-11  \\ 
   38 &   4s$^2$4p$^3$($^2$D)4d$^2$($^3$F)   &  $^4$P$^o _{3/2 }$ &  24.0928 &   3.518-11  \\ 
   39 &   4s$^2$4p$^3$($^2$P)4d$^2$($^3$F)   &  $^4$D$^o _{5/2 }$ &  24.2275 &   2.461-11  \\ 
   40 &   4s$^2$4p$^3$($^2$P)4d$^2$($^3$P)   &  $^4$D$^o _{5/2 }$ &  24.3977 &   1.774-10  \\ 
   41 &   4s$^2$4p$^3$($^4$S)4d$^2$($^3$P)   &  $^6$P$^o _{7/2 }$ &  24.4336 &   5.739-11  \\ 
   42 &   4s$^2$4p$^3$($^4$S)4d$^2$($^3$F)   &  $^6$F$^o _{11/2}$ &  24.5388 &   3.953-08  \\ 
   43 &   4s$^2$4p$^3$($^2$P)4d$^2$($^3$F)   &  $^2$G$^o _{7/2 }$ &  24.9901 &   3.984-12  \\ 
   44 &   4s$^2$4p$^4$($^1$D)4f 	     &  $^2$H$^o _{9/2 }$ &  24.9948 &   7.327-12  \\ 
   45 &   4s$^2$4p$^3$($^2$D)4d$^2$($^3$F)   &  $^4$F$^o _{5/2 }$ &  25.0279 &   3.599-12  \\ 
   46 &   4s$^2$4p$^3$($^2$P)4d$^2$($^1$D)   &  $^2$P$^o _{3/2 }$ &  25.2438 &   3.099-12  \\ 
   47 &   4s$^2$4p$^4$($^1$D)4f 	     &  $^2$D$^o _{3/2 }$ &  25.2599 &   7.771-12  \\ 
   48 &   4s$^2$4p$^3$($^4$S)4d$^2$($^3$F)   &  $^6$F$^o _{1/2 }$ &  25.2688 &   4.826-12  \\ 
   49 &   4s4p$^5$($^3$P)4d		     &  $^4$F$^o _{5/2 }$ &  25.2857 &   1.259-11  \\ 
   50 &   4s$^2$4p$^3$($^2$D)4d$^2$($^1$D)   &  $^2$F$^o _{5/2 }$ &  25.4722 &   8.736-12  \\ 
 \hline            								                	 
\end{tabular}  

\clearpage
\newpage
\begin{tabular}{rllrrrrrrrrrr} \hline
Index  & Configuration                       & Level              & Energy  & $\tau$  (s)   \\
 \hline
   51 &   4s4p$^5$($^3$P)4d		     &  $^4$F$^o _{7/2 }$ &  25.4883 &   6.138-12  \\ 
   52 &   4s$^2$4p$^3$($^4$S)4d$^2$($^3$P)   &  $^4$P$^o _{1/2 }$ &  25.5114 &   4.364-12  \\ 
   53 &   4s$^2$4p$^4$($^1$S)4f 	     &  $^2$F$^o _{5/2 }$ &  25.6214 &   9.985-12  \\ 
   54 &   4s$^2$4p$^3$($^2$P)4d$^2$($^3$F)   &  $^4$F$^o _{3/2 }$ &  25.7739 &   2.222-12  \\ 
   55 &   4s$^2$4p$^3$($^2$P)4d$^2$($^3$F)   &  $^4$F$^o _{9/2 }$ &  25.8007 &   2.075-11  \\ 
   56 &   4s$^2$4p$^3$($^2$D)4d$^2$($^1$G)   &  $^2$I$^o _{11/2}$ &  25.8381 &   3.685-11  \\ 
   57 &   4s$^2$4p$^3$($^2$P)4d$^2$($^1$D)   &  $^2$F$^o _{7/2 }$ &  25.9381 &   2.762-11  \\ 
   58 &   4s4p$^5$($^3$P)4d		     &  $^4$P$^o _{3/2 }$ &  26.0483 &   2.805-12  \\ 
   59 &   4s4p$^5$($^3$P)4d		     &  $^4$D$^o _{5/2 }$ &  26.3546 &   2.217-12  \\ 
   60 &   4s$^2$4p$^4$($^3$P)4d 	     &  $^4$F$   _{3/2 }$ &  26.2847 &   6.222-13  \\ 
   61 &   4s$^2$4p$^4$($^3$P)4f 	     &  $^4$D$^o _{1/2 }$ &  26.4363 &   2.843-12  \\ 
   62 &   4s$^2$4p$^3$($^2$D)4d$^2$($^3$P)   &  $^4$D$^o _{1/2 }$ &  26.7300 &   1.943-12  \\ 
   63 &   4s$^2$4p$^4$($^3$P)4f 	     &  $^2$D$^o _{3/2 }$ &  26.7611 &   9.961-12  \\ 
   64 &   4s$^2$4p$^4$($^3$P)4f 	     &  $^2$F$^o _{7/2 }$ &  26.7815 &   3.258-12  \\ 
   65 &   4s$^2$4p$^4$($^3$P)4f 	     &  $^4$F$^o _{9/2 }$ &  26.8418 &   1.687-12  \\ 
   66 &   4s$^2$4p$^4$($^1$S)4d 	     &  $^2$D$   _{5/2 }$ &  26.8956 &   4.372-10  \\ 
   67 &   4s4p$^5$($^3$P)4d		     &  $^4$P$^o _{5/2 }$ &  27.1442 &   3.084-12  \\ 
   68 &   4s$^2$4p$^4$($^3$P)4f 	     &  $^4$F$^o _{7/2 }$ &  27.2070 &   1.151-12  \\ 
   69 &   4s$^2$4p$^4$($^3$P)4f 	     &  $^4$G$^o _{11/2}$ &  27.2070 &   2.458-12  \\ 
   70 &   4s$^2$4p$^3$($^2$D)4d$^2$($^1$S)   &  $^2$D$^o _{3/2 }$ &  27.2154 &   3.049-12  \\ 
   71 &   4s$^2$4p$^3$($^4$S)4d$^2$($^1$D)   &  $^4$D$^o _{1/2 }$ &  27.2956 &   1.246-12  \\ 
   72 &   4s$^2$4p$^4$($^3$P)4f 	     &  $^4$D$^o _{5/2 }$ &  27.3080 &   1.209-12  \\ 
   73 &   4s$^2$4p$^4$($^3$P)4f 	     &  $^2$G$^o _{9/2 }$ &  27.3660 &   1.250-12  \\ 
   74 &   4s4p$^5$($^3$P)4d		     &  $^4$F$^o _{9/2 }$ &  27.4088 &   2.004-12  \\ 
   75 &   4s4p$^5$($^3$P)4d		     &  $^4$D$^o _{7/2 }$ &  27.4358 &   2.953-12  \\ 
   76 &   4s4p$^5$($^3$P)4d		     &  $^4$F$^o _{3/2 }$ &  27.4611 &   1.206-12  \\ 
   77 &   4s4p$^5$($^3$P)4d		     &  $^2$D$^o _{5/2 }$ &  27.6888 &   3.862-12  \\ 
   78 &   4s4p$^5$($^1$P)4d                  &  $^2$F$^o _{5/2 }$ &  27.8081 &   1.012-12  \\ 
   79 &   4s4p$^5$($^3$P)4d		     &  $^2$F$^o _{7/2 }$ &  27.8172 &   3.989-12  \\ 
   80 &   4s$^2$4p$^3$($^4$S)4d$^2$($^1$D)   &  $^4$D$^o _{3/2 }$ &  27.8684 &   1.664-12  \\ 
   81 &   4s4p$^5$($^1$P)4d		     &  $^2$F$^o _{7/2 }$ &  28.2852 &   1.876-12  \\ 
   82 &   4s$^2$4p$^4$($^3$P)4f 	     &  $^2$D$^o _{5/2 }$ &  28.3289 &   1.442-12  \\ 
   83 &   4s4p$^5$($^3$P)4d		     &  $^2$D$^o _{3/2 }$ &  28.3361 &   1.288-12  \\ 
   84 &   4s$^2$4p$^3$($^4$S)4d$^2$($^1$G)   &  $^4$G$^o _{7/2 }$ &  28.7019 &   9.732-13  \\ 
   85 &   4s$^2$4p$^3$($^4$S)4d$^2$($^1$G)   &  $^4$G$^o _{5/2 }$ &  28.7824 &   1.350-12  \\ 
   86 &   4s4p$^5$($^3$P)4d		     &  $^2$P$^o _{1/2 }$ &  29.1819 &   8.802-13  \\ 
   87 &   4s4p$^5$($^1$P)4d		     &  $^2$D$^o _{3/2 }$ &  29.3774 &   8.999-13  \\ 
   88 &   4s4p$^5$($^3$P)4d		     &  $^4$D$^o _{1/2 }$ &  29.5049 &   6.931-13  \\ 
   89 &   4s4p$^5$($^3$P)4d		     &  $^4$D$^o _{3/2 }$ &  29.5002 &   8.104-13  \\ 
   90 &   4s$^2$4p$^3$($^2$D)4d$^2$($^3$F)   &  $^4$G$^o _{5/2 }$ &  29.5211 &   9.909-13  \\ 
   91 &   4s$^2$4p$^3$($^2$D)4d$^2$($^3$F)   &  $^4$G$^o _{7/2 }$ &  29.7172 &   1.044-12  \\ 
   92 &   4s$^2$4p$^3$($^2$D)4d$^2$($^3$F)   &  $^4$G$^o _{9/2 }$ &  29.9721 &   1.326-12  \\ 
   93 &   4s$^2$4p$^3$($^2$D)4d$^2$($^3$P)   &  $^4$P$^o _{3/2 }$ &  30.2778 &   4.493-12  \\ 
   94 &   4s$^2$4p$^3$($^2$D)4d$^2$($^3$P)   &  $^4$P$^o _{5/2 }$ &  30.2981 &   4.743-12  \\ 
   95 &   4s$^2$4p$^3$($^2$P)4d$^2$($^3$F)   &  $^4$G$^o _{5/2 }$ &  30.3400 &   1.148-12  \\ 
   96 &   4s$^2$4p$^3$($^4$S)4d$^2$($^3$F)   &  $^6$F$^o _{7/2 }$ &  30.3322 &   7.388-12  \\ 
   97 &   4s$^2$4p$^3$($^4$S)4d$^2$($^3$F)   &  $^6$F$^o _{9/2 }$ &  30.3904 &   1.623-11  \\ 
   98 &   4s$^2$4p$^3$($^4$S)4d$^2$($^3$P)   &  $^6$P$^o _{5/2 }$ &  30.4657 &   2.926-11  \\ 
   99 &   4s$^2$4p$^3$($^2$D)4d$^2$($^3$P)   &  $^4$F$^o _{7/2 }$ &  30.5708 &   3.148-11  \\ 
  100 &   4s$^2$4p$^3$($^4$S)4d$^2$($^3$P)   &  $^4$P$^o _{3/2 }$ &  30.6237 &   7.656-13  \\ 
\hline            								                	 
\end{tabular}  

\clearpage
\newpage
\begin{tabular}{rllrrrrrrrrrr} \hline
Index  & Configuration                       & Level              & Energy  & $\tau$  (s)   \\
 \hline
  101 &   4s$^2$4p$^3$($^2$D)4d$^2$($^3$F)   &  $^4$F$^o _{3/2 }$ &  30.6638 &   3.488-12  \\ 
  102 &   4s$^2$4p$^3$($^2$D)4d$^2$($^1$G)   &  $^2$F$^o _{5/2 }$ &  30.6715 &   6.051-12  \\ 
  103 &   4s$^2$4p$^3$($^2$D)4d$^2$($^3$F)   &  $^4$H$^o _{11/2}$ &  30.7200 &   1.403-10  \\ 
  104 &   4s$^2$4p$^3$($^2$D)4d$^2$($^3$P)   &  $^4$F$^o _{9/2 }$ &  30.7486 &   1.461-11  \\ 
  105 &   4s$^2$4p$^3$($^2$D)4d$^2$($^3$P)   &  $^4$D$^o _{5/2 }$ &  30.9644 &   7.656-13  \\ 
  106 &   4s$^2$4p$^3$($^2$D)4d$^2$($^3$F)   &  $^4$P$^o _{1/2 }$ &  31.0257 &   3.548-13  \\ 
  107 &   4s$^2$4p$^3$($^2$D)4d$^2$($^1$G)   &  $^2$H$^o _{11/2}$ &  31.0384 &   1.748-10  \\ 
  108 &   4s$^2$4p$^3$($^2$D)4d$^2$($^3$F)   &  $^4$D$^o _{3/2 }$ &  31.1739 &   3.740-13  \\ 
  109 &   4s$^2$4p$^3$($^2$D)4d$^2$($^3$F)   &  $^4$D$^o _{5/2 }$ &  31.2920 &   3.609-13  \\ 
  110 &   4s$^2$4p$^3$($^2$D)4d$^2$($^1$D)   &  $^2$F$^o _{7/2 }$ &  31.3150 &   3.785-13  \\ 
  111 &   4s$^2$4p$^3$($^2$D)4d$^2$($^1$G)   &  $^2$I$^o _{13/2}$ &  31.2886 &   7.837-07  \\ 
  112 &   4s$^2$4p$^3$($^2$P)4d$^2$($^3$F)   &  $^4$G$^o _{7/2 }$ &  31.4697 &   9.805-12  \\ 
  113 &   4s$^2$4p$^3$($^2$D)4d$^2$($^3$F)   &  $^4$D$^o _{1/2 }$ &  31.5175 &   7.772-13  \\ 
  114 &   4s$^2$4p$^3$($^4$S)4d$^2$($^3$F)   &  $^4$F$^o _{3/2 }$ &  31.5670 &   2.035-12  \\ 
  115 &   4s$^2$4p$^3$($^2$P)4d$^2$($^3$F)   &  $^4$F$^o _{5/2 }$ &  31.5925 &   3.505-12  \\ 
  116 &   4s$^2$4p$^3$($^2$P)4d$^2$($^3$P)   &  $^4$D$^o _{3/2 }$ &  31.6625 &   5.767-12  \\ 
  117 &   4s$^2$4p$^3$($^4$S)4d$^2$($^3$F)   &  $^4$F$^o _{5/2 }$ &  31.6849 &   4.289-12  \\ 
  118 &   4s$^2$4p$^3$($^2$D)4d$^2$($^3$F)   &  $^4$G$^o _{11/2}$ &  31.6619 &   4.508-12  \\ 
  119 &   4s$^2$4p$^3$($^2$P)4d$^2$($^3$P)   &  $^4$P$^o _{1/2 }$ &  31.7827 &   9.973-12  \\ 
  120 &   4s$^2$4p$^3$($^2$D)4d$^2$($^1$G)   &  $^2$G$^o _{7/2 }$ &  31.8142 &   2.127-12  \\ 
  121 &   4s$^2$4p$^3$($^2$P)4d$^2$($^1$G)   &  $^2$H$^o _{9/2 }$ &  31.8419 &   1.851-12  \\ 
  122 &   4s$^2$4p$^3$($^4$S)4d$^2$($^1$D)   &  $^4$D$^o _{7/2 }$ &  31.8587 &   4.434-12  \\ 
  123 &   4s$^2$4p$^3$($^2$D)4d$^2$($^3$F)   &  $^2$H$^o _{9/2 }$ &  31.9066 &   2.598-12  \\ 
  124 &   4s$^2$4p$^3$($^2$P)4d$^2$($^3$P)   &  $^4$D$^o _{1/2 }$ &  31.9686 &   8.495-13  \\ 
  125 &   4s$^2$4p$^3$($^2$P)4d$^2$($^1$D)   &  $^2$D$^o _{3/2 }$ &  31.9999 &   3.438-13  \\ 
  126 &   4s$^2$4p$^3$($^2$D)4d$^2$($^1$D)   &  $^2$P$^o _{3/2 }$ &  32.0297 &   9.743-12  \\ 
  127 &   4s$^2$4p$^3$($^4$S)4d$^2$($^3$F)   &  $^4$F$^o _{9/2 }$ &  32.1176 &   6.846-13  \\ 
  128 &   4s$^2$4p$^3$($^2$D)4d$^2$($^1$D)   &  $^2$D$^o _{3/2 }$ &  32.1400 &   1.862-12  \\ 
  129 &   4s$^2$4p$^3$($^2$D)4d$^2$($^3$F)   &  $^2$G$^o _{7/2 }$ &  32.1537 &   8.324-13  \\ 
  130 &   4s$^2$4p$^3$($^4$S)4d$^2$($^3$P)   &  $^2$P$^o _{1/2 }$ &  32.2018 &   5.667-13  \\ 
  131 &   4s$^2$4p$^3$($^2$D)4d$^2$($^3$F)   &  $^4$F$^o _{9/2 }$ &  32.1955 &   1.344-12  \\ 
  132 &   4s$^2$4p$^3$($^2$D)4d$^2$($^3$P)   &  $^2$D$^o _{5/2 }$ &  32.2157 &   1.423-12  \\ 
  133 &   4s$^2$4p$^3$($^2$D)4d$^2$($^1$G)   &  $^2$D$^o _{3/2 }$ &  32.2322 &   5.690-13  \\ 
  134 &   4s$^2$4p$^3$($^2$D)4d$^2$($^3$P)   &  $^2$F$^o _{7/2 }$ &  32.2737 &   8.840-13  \\ 
  135 &   4s$^2$4p$^3$($^2$D)4d$^2$($^3$F)   &  $^4$P$^o _{5/2 }$ &  32.2854 &   6.761-13  \\ 
  136 &   4s$^2$4p$^3$($^4$S)4d$^2$($^1$G)   &  $^4$G$^o _{11/2}$ &  32.3397 &   6.554-13  \\ 
  137 &   4s$^2$4p$^3$($^2$D)4d$^2$($^1$D)   &  $^2$D$^o _{5/2 }$ &  32.3545 &   1.091-11  \\ 
  138 &   4s$^2$4p$^3$($^2$D)4d$^2$($^1$D)   &  $^2$G$^o _{9/2 }$ &  32.3781 &   1.432-12  \\ 
  139 &   4s$^2$4p$^3$($^2$D)4d$^2$($^3$P)   &  $^4$D$^o _{7/2 }$ &  32.4195 &   5.067-13  \\ 
  140 &   4s$^2$4p$^3$($^2$D)4d$^2$($^3$F)   &  $^4$H$^o _{13/2}$ &  32.4627 &   7.531-07  \\ 
  141 &   4s$^2$4p$^3$($^2$D)4d$^2$($^3$P)   &  $^2$F$^o _{5/2 }$ &  32.5074 &   1.414-12  \\ 
  142 &   4s$^2$4p$^3$($^2$D)4d$^2$($^1$G)   &  $^2$G$^o _{9/2 }$ &  32.5296 &   1.873-12  \\ 
  143 &   4s$^2$4p$^3$($^2$D)4d$^2$($^1$D)   &  $^2$S$^o _{1/2 }$ &  32.6525 &   7.939-13  \\ 
  144 &   4s$^2$4p$^3$($^2$D)4d$^2$($^3$F)   &  $^2$F$^o _{7/2 }$ &  32.6311 &   7.505-12  \\ 
  145 &   4s$^2$4p$^3$($^2$D)4d$^2$($^3$F)   &  $^2$H$^o _{11/2}$ &  32.8560 &   5.634-12  \\ 
  146 &   4s$^2$4p$^3$($^2$D)4d$^2$($^3$P)   &  $^4$D$^o _{3/2 }$ &  32.9126 &   8.632-13  \\ 
  147 &   4s$^2$4p$^3$($^2$P)4d$^2$($^3$F)   &  $^4$D$^o _{7/2 }$ &  32.9583 &   1.110-12  \\ 
  148 &   4s$^2$4p$^4$($^1$D)4f 	     &  $^2$P$^o _{1/2 }$ &  33.0138 &   9.666-13  \\ 
  149 &   4s$^2$4p$^3$($^2$D)4d$^2$($^1$G)   &  $^2$F$^o _{7/2 }$ &  33.0228 &   8.045-13  \\ 
  150 &   4s$^2$4p$^4$($^3$P)4f 	     &  $^4$D$^o _{3/2 }$ &  33.0475 &   9.569-13  \\ 
 \hline            								                	 
\end{tabular}  

\clearpage
\newpage
\begin{tabular}{rllrrrrrrrrrr} \hline
Index  & Configuration                       & Level              & Energy  & $\tau$  (s)   \\
 \hline
  151 &   4s$^2$4p$^4$($^3$P)4f 	     &  $^4$F$^o _{5/2 }$ &  33.0857 &   8.277-13  \\ 
  152 &   4s$^2$4p$^3$($^2$D)4d$^2$($^1$G)   &  $^2$D$^o _{5/2 }$ &  33.1122 &   7.518-13  \\ 
  153 &   4s$^2$4p$^3$($^2$D)4d$^2$($^3$F)   &  $^2$F$^o _{5/2 }$ &  33.2444 &   4.622-13  \\ 
  154 &   4s$^2$4p$^3$($^2$D)4d$^2$($^3$F)   &  $^2$D$^o _{3/2 }$ &  33.2493 &   6.379-13  \\ 
  155 &   4s$^2$4p$^3$($^2$P)4d$^2$($^1$G)   &  $^2$G$^o _{7/2 }$ &  33.3410 &   9.530-13  \\ 
  156 &   4s$^2$4p$^3$($^4$S)4d$^2$($^1$S)   &  $^4$S$^o _{3/2 }$ &  33.4244 &   4.329-12  \\ 
  157 &   4s$^2$4p$^3$($^2$P)4d$^2$($^1$G)   &  $^2$G$^o _{9/2 }$ &  33.5108 &   3.935-11  \\ 
  158 &   4s$^2$4p$^3$($^2$D)4d$^2$($^1$D)   &  $^2$P$^o _{1/2 }$ &  33.5491 &   4.475-13  \\ 
  159 &   4s$^2$4p$^3$($^4$S)4d$^2$($^3$P)   &  $^4$P$^o _{5/2 }$ &  33.6177 &   5.595-13  \\ 
  160 &   4s$^2$4p$^3$($^2$P)4d$^2$($^1$D)   &  $^2$F$^o _{5/2 }$ &  33.6898 &   4.204-12  \\ 
  161 &   4s$^2$4p$^4$($^1$D)4f 	     &  $^2$P$^o _{3/2 }$ &  33.7429 &   5.540-13  \\ 
  162 &   4s$^2$4p$^3$($^2$D)4d$^2$($^3$P)   &  $^4$P$^o _{1/2 }$ &  33.7999 &   1.418-12  \\ 
  163 &   4s$^2$4p$^3$($^4$S)4d$^2$($^1$D)   &  $^4$D$^o _{5/2 }$ &  33.8166 &   1.938-12  \\ 
  164 &   4s$^2$4p$^4$($^1$D)4f 	     &  $^2$F$^o _{7/2 }$ &  33.8490 &   1.137-12  \\ 
  165 &   4s$^2$4p$^3$($^2$D)4d$^2$($^3$F)   &  $^2$G$^o _{9/2 }$ &  33.8509 &   1.534-12  \\ 
  166 &   4s$^2$4p$^3$($^2$D)4d$^2$($^3$P)   &  $^2$P$^o _{3/2 }$ &  33.9405 &   1.079-12  \\ 
  167 &   4s$^2$4p$^3$($^2$D)4d$^2$($^3$P)   &  $^4$F$^o _{3/2 }$ &  33.9611 &   5.367-13  \\ 
  168 &   4s$^2$4p$^4$($^3$P)4f 	     &  $^2$F$^o _{5/2 }$ &  34.0812 &   1.254-12  \\ 
  169 &   4s$^2$4p$^4$($^3$P)4f 	     &  $^4$G$^o _{7/2 }$ &  34.1138 &   1.191-12  \\ 
  170 &   4s$^2$4p$^3$($^2$P)4d$^2$($^3$P)   &  $^4$P$^o _{3/2 }$ &  34.1551 &   5.995-13  \\ 
  171 &   4s$^2$4p$^4$($^3$P)4f 	     &  $^4$F$^o _{3/2 }$ &  34.2595 &   5.235-13  \\ 
  172 &   4s$^2$4p$^3$($^2$D)4d$^2$($^1$S)   &  $^2$D$^o _{5/2 }$ &  34.2536 &   1.095-12  \\ 
  173 &   4s$^2$4p$^3$($^2$P)4d$^2$($^1$S)   &  $^2$P$^o _{1/2 }$ &  34.2895 &   1.954-11  \\ 
  174 &   4s$^2$4p$^4$($^3$P)4f 	     &  $^2$G$^o _{7/2 }$ &  34.3142 &   1.032-12  \\ 
  175 &   4s$^2$4p$^4$($^1$D)4f 	     &  $^2$H$^o _{11/2}$ &  34.3375 &   1.654-12  \\ 
  176 &   4s$^2$4p$^4$($^3$P)4f 	     &  $^4$G$^o _{9/2 }$ &  34.3583 &   7.295-13  \\ 
  177 &   4s$^2$4p$^4$($^3$P)4f 	     &  $^4$G$^o _{5/2 }$ &  34.4085 &   6.496-13  \\ 
  178 &   4s$^2$4p$^3$($^2$D)4d$^2$($^3$P)   &  $^2$D$^o _{3/2 }$ &  34.4506 &   7.855-13  \\ 
  179 &   4s$^2$4p$^3$($^2$D)4d$^2$($^1$G)   &  $^2$H$^o _{9/2 }$ &  34.5211 &   7.813-13  \\ 
  180 &   4s$^2$4p$^4$($^1$D)4f 	     &  $^2$G$^o _{7/2 }$ &  34.5183 &   9.573-13  \\ 
  181 &   4s$^2$4p$^3$($^2$D)4d$^2$($^3$F)   &  $^2$D$^o _{5/2 }$ &  34.6437 &   7.712-13  \\ 
  182 &   4s4p$^5$($^3$P)4d		     &  $^2$F$^o _{5/2 }$ &  34.6813 &   1.497-12  \\ 
  183 &   4s$^2$4p$^2$($^3$P)4d$^3$($^4$F)   &  $^6$G$   _{3/2 }$ &  34.7989 &   1.102-11  \\ 
  184 &   4s$^2$4p$^4$($^1$D)4f 	     &  $^2$G$^o _{9/2 }$ &  34.7912 &   1.759-12  \\ 
  185 &   4s$^2$4p$^4$($^3$P)4f 	     &  $^4$D$^o _{7/2 }$ &  34.8652 &   8.329-13  \\ 
  186 &   4s$^2$4p$^3$($^2$P)4d$^2$($^1$G)   &  $^2$F$^o _{7/2 }$ &  35.0611 &   4.362-13  \\ 
  187 &   4s$^2$4p$^4$($^1$D)4f 	     &  $^2$F$^o _{5/2 }$ &  35.1062 &   4.703-13  \\ 
  188 &   4s4p$^5$($^1$P)4d		     &  $^2$P$^o _{3/2 }$ &  35.2229 &   7.874-13  \\ 
  189 &   4s4p$^5$($^1$P)4d		     &  $^2$P$^o _{1/2 }$ &  35.4454 &   3.586-13  \\ 
  190 &   4s4p$^5$($^1$P)4d		     &  $^2$D$^o _{5/2 }$ &  35.4299 &   7.846-13  \\ 
  191 &   4s4p$^5$($^3$P)4d		     &  $^2$P$^o _{3/2 }$ &  35.4765 &   3.672-13  \\ 
  192 &   4s$^2$4p$^3$($^2$D)4d$^2$($^3$F)   &  $^4$D$^o _{7/2 }$ &  35.4617 &   7.643-13  \\ 
  193 &   4s$^2$4p$^3$($^2$D)4d$^2$($^3$P)   &  $^2$P$^o _{1/2 }$ &  35.9327 &   5.424-13  \\ 
  194 &   4s$^2$4p$^4$($^1$D)4f 	     &  $^2$D$^o _{5/2 }$ &  35.9782 &   6.625-13  \\ 
  195 &   4s$^2$4p$^2$($^3$P)4d$^3$($^4$F)   &  $^6$G$   _{5/2 }$ &  36.2982 &   5.322-12  \\ 
  196 &   4s$^2$4p$^2$($^3$P)4d$^3$($^2$H)   &  $^4$I$   _{9/2 }$ &  36.4768 &   7.931-12  \\ 
  197 &   4s$^2$4p$^2$($^3$P)4d$^3$($^2$G)   &  $^4$H$   _{7/2 }$ &  36.5089 &   6.733-12  \\ 
  198 &   4s$^2$4p$^2$($^3$P)4d$^3$($^4$P)   &  $^6$D$   _{3/2 }$ &  36.5322 &   4.238-12  \\ 
  199 &   4s$^2$4p$^3$($^2$D)4d$^2$($^3$F)   &  $^2$P$^o _{3/2 }$ &  36.5867 &   4.492-13  \\ 
  200 &   4s$^2$4p$^2$($^3$P)4d$^3$($^4$P)   &  $^6$D$   _{1/2 }$ &  36.6707 &   4.195-12  \\ 
 \hline            								                	 
\end{tabular}  

\clearpage
\newpage
\begin{tabular}{rllrrrrrrrrrr} \hline
Index  & Configuration                       & Level              & Energy  & $\tau$  (s)   \\
 \hline
  201 &   4s$^2$4p$^2$($^3$P)4d$^3$($^2$F)   &  $^4$G$   _{5/2 }$ &  37.2564 &   6.044-12  \\ 
  202 &   4s$^2$4p$^3$($^2$P)4d[$^3$F]4f     &  $^4$I$   _{9/2 }$ &  37.4411 &   6.450-11  \\ 
  203 &   4s$^2$4p$^3$($^4$S)4d[$^5$D]4f     &  $^6$H$   _{7/2 }$ &  37.5443 &   1.366-11  \\ 
  204 &   4s$^2$4p$^3$($^2$P)4d[$^3$F]4f     &  $^4$I$   _{11/2}$ &  37.7860 &   3.817-11  \\ 
  205 &   4s$^2$4p$^3$($^4$S)4d[$^5$D]4f     &  $^6$F$   _{1/2 }$ &  37.8422 &   4.798-12  \\ 
  206 &   4s$^2$4p$^3$($^4$S)4d[$^5$D]4f     &  $^6$F$   _{3/2 }$ &  37.8573 &   4.084-12  \\ 
  207 &   4s$^2$4p$^3$($^4$S)4d[$^5$D]4f     &  $^6$H$   _{5/2 }$ &  37.8636 &   1.825-11  \\ 
  208 &   4s$^2$4p$^2$($^1$S)4d$^3$($^4$F)   &  $^4$F$   _{7/2 }$ &  37.8375 &   6.831-12  \\ 
  209 &   4s$^2$4p$^3$($^4$S)4d[$^5$D]4f     &  $^6$F$   _{5/2 }$ &  37.9573 &   3.662-12  \\ 
  210 &   4s$^2$4p$^2$($^1$S)4d$^3$($^4$F)   &  $^4$F$   _{9/2 }$ &  38.0020 &   6.820-12  \\ 
  211 &   4s$^2$4p$^3$($^4$S)4d[$^5$D]4f     &  $^4$D$   _{7/2 }$ &  38.0310 &   4.664-12  \\ 
  212 &   4s$^2$4p$^3$($^4$S)4d[$^5$D]4f     &  $^6$H$   _{11/2}$ &  38.0908 &   1.398-11  \\ 
  213 &   4s$^2$4p$^2$($^1$S)4d$^3$($^4$P)   &  $^4$P$   _{5/2 }$ &  38.1382 &   6.628-12  \\ 
  214 &   4s$^2$4p$^3$($^4$S)4d[$^5$D]4f     &  $^6$F$   _{11/2}$ &  38.2583 &   5.915-12  \\ 
  215 &   4s$^2$4p$^2$($^3$P)4d$^3$($^2$D)   &  $^4$F$   _{3/2 }$ &  38.2945 &   3.624-12  \\ 
  216 &   4s$^2$4p$^3$($^2$P)4d[$^3$D]4f     &  $^4$G$   _{9/2 }$ &  38.3513 &   4.391-12  \\ 
  217 &   4s$^2$4p$^3$($^2$P)4d$^2$($^3$F)   &  $^4$D$^o _{3/2 }$ &  38.3066 &   2.726-13  \\ 
  218 &   4s$^2$4p$^3$($^2$D)4d$^2$($^3$F)   &  $^2$P$^o _{1/2 }$ &  38.3192 &   2.568-13  \\ 
  219 &   4s$^2$4p$^3$($^2$P)4d$^2$($^1$D)   &  $^2$D$^o _{5/2 }$ &  38.3313 &   3.154-13  \\ 
  220 &   4s$^2$4p$^3$($^4$S)4d[$^5$D]4f     &  $^6$F$   _{7/2 }$ &  38.4400 &   2.294-12  \\ 
  221 &   4s$^2$4p$^3$($^4$S)4d[$^5$D]4f     &  $^4$G$   _{5/2 }$ &  38.4433 &   4.688-12  \\ 
  222 &   4s$^2$4p$^2$($^3$P)4d$^3$($^2$P)   &  $^4$D$   _{1/2 }$ &  38.4503 &   2.946-12  \\ 
  223 &   4s$^2$4p$^3$($^4$S)4d[$^5$D]4f     &  $^4$G$   _{7/2 }$ &  38.4693 &   5.898-12  \\ 
  224 &   4s$^2$4p$^3$($^2$P)4d[$^3$F]4f     &  $^4$I$   _{13/2}$ &  38.5208 &   1.780-11  \\ 
  225 &   4s$^2$4p$^3$($^2$P)4d[$^3$D]4f     &  $^4$F$   _{3/2 }$ &  38.5572 &   1.930-12  \\ 
  226 &   4s$^2$4p$^3$($^4$S)4d[$^5$D]4f     &  $^6$G$   _{5/2 }$ &  38.6031 &   1.216-12  \\ 
  227 &   4s$^2$4p$^3$($^4$S)4d$^2$($^3$F)   &  $^4$D$^o _{7/2 }$ &  38.5104 &   3.105-13  \\ 
  228 &   4s$^2$4p$^3$($^2$P)4d[$^3$P]4f     &  $^4$G$   _{9/2 }$ &  38.5901 &   4.341-12  \\ 
  229 &   4s$^2$4p$^3$($^2$P)4d[$^3$F]4f     &  $^4$G$   _{7/2 }$ &  38.7020 &   1.598-12  \\ 
  230 &   4s$^2$4p$^3$($^2$P)4d$^2$($^3$F)   &  $^2$F$^o _{5/2 }$ &  38.6603 &   4.416-13  \\ 
  231 &   4s$^2$4p$^3$($^2$P)4d[$^3$F]4f     &  $^4$G$   _{9/2 }$ &  38.7616 &   1.732-12  \\ 
  232 &   4s$^2$4p$^3$($^4$S)4d[$^5$D]4f     &  $^6$G$   _{3/2 }$ &  38.8038 &   1.589-12  \\ 
  233 &   4s$^2$4p$^3$($^4$S)4d[$^3$D]4f     &  $^2$D$   _{5/2 }$ &  38.7864 &   2.338-12  \\ 
  234 &   4s$^2$4p$^3$($^2$D)4d$^2$($^3$F)   &  $^4$F$^o _{7/2 }$ &  38.7299 &   5.175-13  \\ 
  235 &   4s$^2$4p$^3$($^2$D)4d$^2$($^3$F)   &  $^4$H$^o _{9/2 }$ &  38.7267 &   6.001-13  \\ 
  236 &   4s$^2$4p$^3$($^2$P)4d[$^3$P]4f     &  $^4$D$   _{7/2 }$ &  38.8646 &   3.297-12  \\ 
  237 &   4s$^2$4p$^3$($^2$P)4d$^2$($^3$P)   &  $^4$S$^o _{3/2 }$ &  38.8715 &   4.126-13  \\ 
  238 &   4s$^2$4p$^3$($^2$P)4d$^2$($^3$P)   &  $^4$P$^o _{5/2 }$ &  38.8800 &   5.550-13  \\ 
  239 &   4s$^2$4p$^3$($^2$P)4d$^2$($^3$F)   &  $^4$G$^o _{11/2}$ &  38.8808 &   8.715-13  \\ 
  240 &   4s$^2$4p$^3$($^4$S)4d[$^5$D]4f     &  $^6$D$   _{3/2 }$ &  39.0236 &   1.688-12  \\ 
  241 &   4s$^2$4p$^3$($^4$S)4d$^2$($^3$P)   &  $^2$P$^o _{3/2 }$ &  38.9353 &   3.453-13  \\ 
  242 &   4s$^2$4p$^3$($^4$S)4d[$^5$D]4f     &  $^6$P$   _{5/2 }$ &  39.0436 &   1.543-12  \\ 
  243 &   4s$^2$4p$^3$($^4$S)4d[$^5$D]4f     &  $^6$D$   _{1/2 }$ &  39.1339 &   1.635-12  \\ 
  244 &   4s$^2$4p$^3$($^4$S)4d$^2$($^3$F)   &  $^2$F$^o _{7/2 }$ &  39.0268 &   6.102-13  \\ 
  245 &   4s$^2$4p$^3$($^2$P)4d$^2$($^3$P)   &  $^2$S$^o _{1/2 }$ &  39.0768 &   6.573-13  \\ 
  246 &   4s$^2$4p$^3$($^2$P)4d[$^3$D]4f     &  $^2$F$   _{5/2 }$ &  39.2277 &   3.401-12  \\ 
  247 &   4s$^2$4p$^3$($^2$D)4d[$^3$G]4f     &  $^4$K$   _{11/2}$ &  39.3029 &   2.971-12  \\ 
  248 &   4s$^2$4p$^3$($^2$P)4d[$^3$P]4f     &  $^4$G$   _{7/2 }$ &  39.3671 &   2.568-12  \\ 
  249 &   4s$^2$4p$^3$($^4$S)4d[$^5$D]4f     &  $^6$F$   _{9/2 }$ &  39.3325 &   8.911-12  \\ 
  250 &   4s$^2$4p$^3$($^2$D)4d$^2$($^1$D)   &  $^2$G$^o _{7/2 }$ &  39.2706 &   7.968-13  \\ 
 \hline            								                	 
\end{tabular}  

\clearpage
\newpage
\begin{tabular}{rllrrrrrrrrrr} \hline
Index  & Configuration                       & Level              & Energy  & $\tau$  (s)   \\
 \hline
  251 &   4s$^2$4p$^3$($^2$D)4d$^2$($^3$P)   &  $^4$F$^o _{5/2 }$ &  39.3037 &   1.243-12  \\ 
  252 &   4s$^2$4p$^3$($^4$S)4d[$^5$D]4f     &  $^4$G$   _{9/2 }$ &  39.4118 &   1.178-12  \\ 
  253 &   4s$^2$4p$^2$($^3$P)4d$^3$($^2$D)   &  $^4$F$   _{3/2 }$ &  39.4304 &   2.623-12  \\ 
  254 &   4s$^2$4p$^3$($^2$P)4d$^2$($^3$F)   &  $^2$G$^o _{9/2 }$ &  39.3570 &   1.219-12  \\ 
  255 &   4s$^2$4p$^3$($^2$P)4d$^2$($^1$G)   &  $^2$H$^o _{11/2}$ &  39.3478 &   2.147-12  \\ 
  256 &   4s$^2$4p$^3$($^2$P)4d[$^3$P]4f     &  $^4$D$   _{1/2 }$ &  39.4918 &   2.068-12  \\ 
  257 &   4s$^2$4p$^3$($^2$P)4d$^2$($^3$P)   &  $^2$D$^o _{3/2 }$ &  39.4184 &   6.135-13  \\ 
  258 &   4s$^2$4p$^3$($^2$P)4d[$^3$P]4f     &  $^2$G$   _{7/2 }$ &  39.5145 &   1.903-12  \\ 
  259 &   4s$^2$4p$^3$($^2$P)4d[$^3$F]4f     &  $^4$I$   _{15/2}$ &  39.4762 &   2.918-07  \\ 
  260 &   4s$^2$4p$^3$($^2$P)4d[$^3$P]4f     &  $^4$G$   _{11/2}$ &  39.5014 &   4.190-12  \\ 
  261 &   4s$^2$4p$^3$($^2$P)4d[$^3$P]4f     &  $^4$F$   _{3/2 }$ &  39.5543 &   1.869-12  \\ 
  262 &   4s$^2$4p$^3$($^2$P)4d[$^3$F]4f     &  $^2$G$   _{9/2 }$ &  39.5642 &   1.429-12  \\ 
  263 &   4s$^2$4p$^3$($^2$P)4d$^2$($^3$P)   &  $^2$D$^o _{5/2 }$ &  39.4699 &   1.112-12  \\ 
  264 &   4s$^2$4p$^3$($^2$P)4d$^2$($^3$P)   &  $^4$D$^o _{7/2 }$ &  39.4742 &   2.381-12  \\ 
  265 &   4s$^2$4p$^3$($^2$P)4d[$^3$F]4f     &  $^4$G$   _{11/2}$ &  39.6604 &   2.247-12  \\ 
  266 &   4s$^2$4p$^3$($^2$P)4d[$^1$D]4f     &  $^2$P$   _{3/2 }$ &  39.6948 &   2.635-12  \\ 
  267 &   4s$^2$4p$^3$($^2$P)4d[$^3$F]4f     &  $^4$F$   _{7/2 }$ &  39.7084 &   2.070-12  \\ 
  268 &   4s$^2$4p$^3$($^4$S)4d[$^5$D]4f     &  $^4$D$   _{1/2 }$ &  39.7984 &   2.276-12  \\ 
  269 &   4s$^2$4p$^3$($^2$P)4d[$^3$F]4f     &  $^4$D$   _{5/2 }$ &  39.8252 &   2.897-12  \\ 
  270 &   4s$^2$4p$^3$($^4$S)4d$^2$($^1$G)   &  $^4$G$^o _{9/2 }$ &  39.7327 &   1.339-12  \\ 
  271 &   4s$^2$4p$^3$($^2$P)4d[$^3$F]4f     &  $^4$H$   _{13/2}$ &  39.8222 &   2.067-12  \\ 
  272 &   4s$^2$4p$^3$($^2$P)4d$^2$($^1$D)   &  $^2$P$^o _{1/2 }$ &  39.7796 &   1.581-11  \\ 
  273 &   4s4p$^5$($^3$P)4f		     &  $^4$F$   _{5/2 }$ &  39.9006 &   1.668-12  \\ 
  274 &   4s$^2$4p$^3$($^4$S)4d$^2$($^3$F)   &  $^2$F$^o _{5/2 }$ &  39.8304 &   8.526-13  \\ 
  275 &   4s$^2$4p$^3$($^2$P)4d[$^3$F]4f     &  $^4$P$   _{3/2 }$ &  39.9339 &   2.968-12  \\ 
  276 &   4s$^2$4p$^3$($^2$P)4d[$^3$F]4f     &  $^4$F$   _{9/2 }$ &  39.9829 &   1.624-12  \\ 
  277 &   4s$^2$4p$^3$($^2$P)4d[$^3$P]4f     &  $^2$F$   _{5/2 }$ &  39.9930 &   2.012-12  \\ 
  278 &   4s4p$^5$($^3$P)4f		     &  $^4$F$   _{7/2 }$ &  40.0810 &   1.643-12  \\ 
  279 &   4s$^2$4p$^3$($^2$P)4d[$^3$P]4f     &  $^2$D$   _{5/2 }$ &  40.1167 &   1.256-12  \\ 
  280 &   4s$^2$4p$^3$($^4$S)4d[$^5$D]4f     &  $^6$P$   _{3/2 }$ &  40.1715 &   2.774-12  \\ 
  281 &   4s$^2$4p$^3$($^2$P)4d[$^3$F]4f     &  $^4$D$   _{7/2 }$ &  40.2315 &   2.670-12  \\ 
  282 &   4s$^2$4p$^3$($^2$P)4d[$^3$F]4f     &  $^2$H$   _{11/2}$ &  40.2448 &   2.491-12  \\ 
  283 &   4s$^2$4p$^3$($^2$P)4d[$^3$F]4f     &  $^4$P$   _{5/2 }$ &  40.2846 &   3.472-12  \\ 
  284 &   4s$^2$4p$^3$($^2$P)4d$^2$($^3$F)   &  $^2$F$^o _{7/2 }$ &  40.2252 &   1.626-11  \\ 
  285 &   4s$^2$4p$^3$($^2$P)4d[$^1$D]4f     &  $^2$G$   _{9/2 }$ &  40.3648 &   1.816-12  \\ 
  286 &   4s$^2$4p$^3$($^2$P)4d[$^3$P]4f     &  $^2$F$   _{7/2 }$ &  40.4118 &   1.823-12  \\ 
  287 &   4s$^2$4p$^3$($^4$S)4d[$^3$D]4f     &  $^4$D$   _{3/2 }$ &  40.6137 &   1.391-12  \\ 
  288 &   4s$^2$4p$^3$($^2$P)4d[$^3$F]4f     &  $^2$I$   _{13/2}$ &  40.6099 &   2.563-12  \\ 
  289 &   4s$^2$4p$^3$($^2$P)4d[$^3$P]4f     &  $^2$G$   _{9/2 }$ &  40.6484 &   1.848-12  \\ 
  290 &   4s$^2$4p$^3$($^2$P)4d$^2$($^1$S)   &  $^2$P$^o _{3/2 }$ &  40.5814 &   3.244-12  \\ 
  291 &   4s$^2$4p$^3$($^2$P)4d[$^1$P]4f     &  $^2$F$   _{5/2 }$ &  40.7483 &   1.288-12  \\ 
  292 &   4s$^2$4p$^3$($^2$P)4d[$^3$D]4f     &  $^2$F$   _{7/2 }$ &  40.7761 &   1.115-12  \\ 
  293 &   4s$^2$4p$^3$($^2$P)4d$^2$($^3$F)   &  $^2$D$^o _{3/2 }$ &  40.7061 &   4.150-13  \\ 
  294 &   4s$^2$4p$^3$($^2$P)4d[$^3$D]4f     &  $^4$F$   _{9/2 }$ &  40.9155 &   9.802-13  \\ 
  295 &   4s$^2$4p$^3$($^2$P)4d[$^1$F]4f     &  $^2$D$   _{3/2 }$ &  40.9885 &   1.277-12  \\ 
  296 &   4s$^2$4p$^3$($^2$P)4d[$^3$D]4f     &  $^4$F$   _{5/2 }$ &  41.0028 &   1.109-12  \\ 
  297 &   4s4p$^5$($^3$P)4f		     &  $^4$D$   _{1/2 }$ &  41.0377 &   1.064-12  \\ 
  298 &   4s$^2$4p$^3$($^2$P)4d[$^3$D]4f     &  $^4$G$   _{11/2}$ &  41.0404 &   1.391-12  \\ 
  299 &   4s$^2$4p$^3$($^2$P)4d[$^1$F]4f     &  $^2$F$   _{7/2 }$ &  41.0763 &   1.522-12  \\ 
  300 &   4s4p$^5$($^3$P)4f		     &  $^4$D$   _{3/2 }$ &  41.1978 &   9.994-13  \\ 
 \hline            								                	 
\end{tabular}  

\clearpage
\newpage
\begin{tabular}{rllrrrrrrrrrr} \hline
Index  & Configuration                       & Level              & Energy  & $\tau$  (s)   \\
 \hline
  301 &   4s$^2$4p$^3$($^2$P)4d[$^1$F]4f     &  $^2$P$   _{1/2 }$ &  41.2134 &   1.027-12  \\ 
  302 &   4s4p$^5$($^1$P)4f		     &  $^2$D$   _{3/2 }$ &  41.2687 &   1.516-12  \\ 
  303 &   4s$^2$4p$^3$($^2$P)4d[$^1$P]4f     &  $^2$G$   _{9/2 }$ &  41.2760 &   2.039-12  \\ 
  304 &   4s$^2$4p$^3$($^2$D)4d[$^3$G]4f     &  $^2$K$   _{13/2}$ &  41.2769 &   1.355-12  \\ 
  305 &   4s$^2$4p$^3$($^4$S)4d[$^5$D]4f     &  $^4$P$   _{1/2 }$ &  41.3689 &   2.310-12  \\ 
  306 &   4s$^2$4p$^3$($^2$P)4d$^2$($^3$P)   &  $^2$P$^o _{1/2 }$ &  41.3253 &   3.110-13  \\ 
  307 &   4s4p$^5$($^3$P)4f		     &  $^4$D$   _{5/2 }$ &  41.4078 &   1.508-12  \\ 
  308 &   4s4p$^5$($^3$P)4f		     &  $^4$G$   _{7/2 }$ &  41.4607 &   1.299-12  \\ 
  309 &   4s4p$^5$($^3$P)4f		     &  $^4$F$   _{3/2 }$ &  41.4655 &   1.604-12  \\ 
  310 &   4s4p$^5$($^3$P)4f		     &  $^4$G$   _{9/2 }$ &  41.4643 &   1.261-12  \\ 
  311 &   4s4p$^5$($^3$P)4f		     &  $^4$G$   _{11/2}$ &  41.4665 &   1.148-12  \\ 
  312 &   4s4p$^5$($^3$P)4f		     &  $^2$F$   _{5/2 }$ &  41.4864 &   1.159-12  \\ 
  313 &   4s$^2$4p$^2$($^3$P)4d$^3$($^2$D)   &  $^4$D$   _{1/2 }$ &  41.5531 &   8.096-13  \\ 
  314 &   4s$^2$4p$^3$($^2$P)4d[$^1$F]4f     &  $^2$D$   _{5/2 }$ &  41.6322 &   1.542-12  \\ 
  315 &   4s4p$^5$($^3$P)4f		     &  $^2$F$   _{7/2 }$ &  41.7308 &   1.194-12  \\ 
  316 &   4s$^2$4p$^3$($^2$P)4d[$^1$F]4f     &  $^2$S$   _{1/2 }$ &  41.7401 &   1.297-12  \\ 
  317 &   4s4p$^5$($^3$P)4f		     &  $^4$D$   _{7/2 }$ &  41.7288 &   1.108-12  \\ 
  318 &   4s$^2$4p$^3$($^2$P)4d$^2$($^3$P)   &  $^2$P$^o _{3/2 }$ &  41.6480 &   7.838-13  \\ 
  319 &   4s$^2$4p$^3$($^2$P)4d[$^3$D]4f     &  $^2$H$   _{11/2}$ &  41.7475 &   8.543-13  \\ 
  320 &   4s4p$^5$($^3$P)4f		     &  $^2$D$   _{3/2 }$ &  41.7663 &   9.165-13  \\ 
  321 &   4s$^2$4p$^2$($^3$P)4d$^3$($^4$F)   &  $^6$F$   _{3/2 }$ &  41.8170 &   7.889-13  \\ 
  322 &   4s4p$^5$($^3$P)4f		     &  $^4$G$   _{5/2 }$ &  41.8470 &   1.226-12  \\ 
  323 &   4s4p$^5$($^3$P)4f		     &  $^2$G$   _{9/2 }$ &  41.9010 &   1.124-12  \\ 
  324 &   4s$^2$4p$^3$($^2$P)4d$^2$($^3$F)   &  $^2$D$^o _{5/2 }$ &  41.8290 &   6.023-13  \\ 
  325 &   4s$^2$4p$^4$($^1$S)4f 	     &  $^2$F$^o _{7/2 }$ &  41.8711 &   8.508-13  \\ 
  326 &   4s$^2$4p$^2$($^3$P)4d$^3$($^4$F)   &  $^6$F$   _{5/2 }$ &  41.9804 &   7.472-13  \\ 
  327 &   4s4p$^5$($^3$P)4f		     &  $^2$G$   _{7/2 }$ &  42.0593 &   9.701-13  \\ 
  328 &   4s$^2$4p$^2$($^1$D)4d$^3$($^4$F)   &  $^4$H$   _{7/2 }$ &  42.1444 &   9.123-13  \\ 
  329 &   4s$^2$4p$^3$($^2$P)4d[$^1$P]4f     &  $^2$D$   _{3/2 }$ &  42.1586 &   1.171-12  \\ 
  330 &   4s$^2$4p$^3$($^2$P)4d$^2$($^1$G)   &  $^2$F$^o _{5/2 }$ &  42.1726 &   3.686-13  \\ 
  331 &   4s4p$^5$($^3$P)4f		     &  $^2$D$   _{5/2 }$ &  42.3479 &   1.142-12  \\ 
  332 &   4s4p$^5$($^3$P)4f		     &  $^4$F$   _{9/2 }$ &  42.4077 &   1.057-12  \\ 
  333 &   4s$^2$4p$^2$($^3$P)4d$^3$($^4$F)   &  $^6$G$   _{7/2 }$ &  42.4498 &   1.401-12  \\ 
  334 &   4s$^2$4p$^2$($^3$P)4d$^3$($^2$P)   &  $^4$D$   _{5/2 }$ &  42.4592 &   1.403-12  \\ 
  335 &   4s$^2$4p$^2$($^3$P)4d$^3$($^4$P)   &  $^6$P$   _{3/2 }$ &  42.4851 &   1.642-12  \\ 
  336 &   4s4p$^5$($^1$P)4f		     &  $^2$G$   _{7/2 }$ &  42.5133 &   1.034-12  \\ 
  337 &   4s$^2$4p$^2$($^3$P)4d$^3$($^4$F)   &  $^6$G$   _{9/2 }$ &  42.5342 &   1.509-12  \\ 
  338 &   4s4p$^5$($^1$P)4f		     &  $^2$F$   _{5/2 }$ &  42.5732 &   1.029-12  \\ 
  339 &   4s$^2$4p$^3$($^4$S)4d[$^3$D]4f     &  $^2$P$   _{3/2 }$ &  42.6916 &   1.033-12  \\ 
  340 &   4s4p$^5$($^1$P)4f		     &  $^2$G$   _{9/2 }$ &  42.7559 &   1.061-12  \\ 
  341 &   4s$^2$4p$^2$($^3$P)4d$^3$($^4$F)   &  $^6$F$   _{1/2 }$ &  42.7647 &   9.289-13  \\ 
  342 &   4s4p$^5$($^1$P)4f		     &  $^2$D$   _{5/2 }$ &  42.7749 &   1.369-12  \\ 
  343 &   4s$^2$4p$^2$($^3$P)4d$^3$($^4$F)   &  $^4$F$   _{3/2 }$ &  42.7873 &   1.075-12  \\ 
  344 &   4s$^2$4p$^2$($^3$P)4d$^3$($^4$F)   &  $^6$G$   _{11/2}$ &  42.8022 &   1.796-12  \\ 
  345 &   4s4p$^5$($^1$P)4f		     &  $^2$F$   _{7/2 }$ &  42.8457 &   1.213-12  \\ 
  346 &   4s$^2$4p$^2$($^3$P)4d$^3$($^4$F)   &  $^4$F$   _{5/2 }$ &  42.9412 &   7.338-13  \\ 
  347 &   4s$^2$4p$^2$($^1$D)4d$^3$($^4$F)   &  $^4$D$   _{1/2 }$ &  42.9632 &   5.222-13  \\ 
  348 &   4s$^2$4p$^2$($^3$P)4d$^3$($^2$G)   &  $^4$G$   _{7/2 }$ &  43.0064 &   1.172-12  \\ 
  349 &   4s$^2$4p$^2$($^1$D)4d$^3$($^2$H)   &  $^2$H$   _{9/2 }$ &  43.0361 &   1.436-12  \\ 
  350 &   4s$^2$4p$^2$($^1$D)4d$^3$($^2$G)   &  $^2$I$   _{13/2}$ &  43.0840 &   1.570-12  \\ 
 \hline            								                	 
\end{tabular}  

\clearpage
\newpage

\begin{tabular}{rllrrrrrrrrrr} \hline
Index  & Configuration                       & Level              & Energy  & $\tau$  (s)   \\
 \hline
  351 &   4s$^2$4p$^2$($^3$P)4d$^3$($^4$F)   &  $^6$D$   _{3/2 }$ &  43.1717 &   4.522-13  \\ 
  352 &   4s$^2$4p$^2$($^3$P)4d$^3$($^4$P)   &  $^6$D$   _{5/2 }$ &  43.2639 &   4.831-13  \\ 
  353 &   4s$^2$4p$^2$($^3$P)4d$^3$($^2$F)   &  $^4$D$   _{7/2 }$ &  43.3970 &   1.437-12  \\ 
  354 &   4s$^2$4p$^2$($^1$D)4d$^3$($^2$G)   &  $^2$H$   _{9/2 }$ &  43.4078 &   1.532-12  \\ 
  355 &   4s$^2$4p$^2$($^3$P)4d$^3$($^2$H)   &  $^4$I$   _{11/2}$ &  43.4029 &   1.581-12  \\ 
  356 &   4p$^6$4d			     &  $^2$D$   _{3/2 }$ &  43.5244 &   5.304-13  \\ 
  357 &   4s$^2$4p$^2$($^3$P)4d$^3$($^4$P)   &  $^6$D$   _{7/2 }$ &  43.4998 &   1.208-12  \\ 
  358 &   4s$^2$4p$^2$($^3$P)4d$^3$($^4$F)   &  $^4$G$   _{5/2 }$ &  43.5981 &   1.021-12  \\ 
  359 &   4s$^2$4p$^2$($^3$P)4d$^3$($^4$F)   &  $^6$F$   _{9/2 }$ &  43.6047 &   2.241-12  \\ 
  360 &   4s$^2$4p$^3$($^2$D)4d[$^3$D]4f     &  $^4$G$   _{5/2 }$ &  43.6906 &   2.902-12  \\
 \hline            								                	 
\end{tabular}   								   					       
			      							   					       

\begin{flushleft}													       
{\small
																														       
}															       
\end{flushleft} 


\begin{thebibliography}{99}

\bibitem{kbf}   K. Fournier.  At.  Data Nucl. Data Tables  68 (1998) 1.
\bibitem{jgk}   V. Jonauskas, G. Gaigalas, S. Ku{\v {c}}as.  At.  Data Nucl. Data Tables  98 (2012) 19.

\bibitem{sbu} S.B. Utter, P. Beiersdorfer,  E. Tr{\" {a}}bert.  Can. J. Phys.  80 (2002)  1503.
\bibitem{ll1}     J. Clementson, P. Beiersdorfer, G.V. Brown, M.F. Gu, H. Lundberg, Y. Podpaly,   E. Tr{\" {a}}bert. Can. J. Phys.  89 (2011) 571.
\bibitem{cjp}   K.M. Aggarwal,  F.P. Keenan.  Can. J. Phys. 92 (2014) xxx,  http://arxiv.org/abs/1310.6620
\bibitem{ajm} S. Aggarwal, A.K.S. Jha,  M. Mohan.  Can. J. Phys. 91 (2013) 394.
\bibitem{grasp0} I.P. Grant, B.J.  McKenzie, P.H. Norrington, D.F. Mayers,   N.C. Pyper. Comput. Phys. Commun.   21  (1980) 207.
\bibitem{grasp2k} P. J\"{o}nsson, X. He, C.F. Fischer,  I.P. Grant.  Comput. Phys. Commun.  177 (2007) 597.
\bibitem{grasp2kk} P. J\"{o}nsson P, G. Gaigalas, J. Biero\'{n}, C.F. Fischer,  I.P. Grant.  Comput. Phys. Commun.  184 (2013) 2197.
\bibitem{kr}    K.M.  Aggarwal, F.P. Keenan,  K.D. Lawson.   At. Data Nucl. Data Tables   94 (2008) 323.
\bibitem{xe}   K.M.  Aggarwal, F.P.  Keenan,  K.D. Lawson.   At. Data Nucl. Data Tables   96 (2010) 123.
\bibitem{tivi}  K.M. Aggarwal,  F.P. Keenan. Phys. Scr. 88 (2013) 025302.
\bibitem{tivii}  K.M. Aggarwal,  F.P. Keenan. Phys. Scr. 88 (2013) 065304.
\bibitem{fac}   M.F.  Gu.  Can. J. Phys.  86 (2008) 675.
\bibitem{tix}  K.M. Aggarwal,  F.P. Keenan. Phys. Scr. 88 (2013) 025303.
\bibitem{fe15}   K.M. Aggarwal, V. Tayal, G.P. Gupta, F.P. Keenan.   At. Data Nucl. Data Tables 93 (2007) 615.
\bibitem{fst}   K.M. Aggarwal,  F.P. Keenan.  Fusion Sci. Tech.  63 (2013) 363.
\bibitem{ah1}  A. Hibbert. Rep. Prog. Phys.  38 (1975) 1217. 
\bibitem{ah2}  P. Oliver, A. Hibbert. AIP Conf. Proc.  1344 (2011) 70.
\bibitem{kma} K.M. Aggarwal.  Astrophys. J. Suppl.  118 (1998) 589.


\end{thebibliography}
\end{document}